\documentclass[11pt]{article}
\usepackage[left=1in,top=1in,right=1in,bottom=1in,head=0in,nofoot]{geometry}

\usepackage{setspace,url,bm,amsmath} 
\usepackage{scalerel}

\usepackage{xcolor}
 \usepackage{multirow}
 \usepackage{arydshln}
 \usepackage{mathtools}
\usepackage{titlesec} 
\titlelabel{\thetitle.\quad} 
\titleformat*{\section}{\bf\Large\center}

\usepackage{authblk}
\usepackage{float}
\usepackage{graphicx} 
\usepackage{bbm}
\usepackage{latexsym}
\usepackage{caption}
\usepackage[margin=20pt]{subcaption}
\usepackage{enumerate}
\graphicspath{ {./figs/} }

\usepackage{amsthm}
\usepackage{amssymb}
\usepackage{amsmath}
\usepackage{color}

\usepackage{comment}
\theoremstyle{definition}
\newtheorem{assumption}{Assumption}
\newtheorem*{theorem*}{Theorem}
\newtheorem{theorem}{Theorem}
\newtheorem*{rmk*}{remark}
\newtheorem{proposition}{Proposition}

\newtheorem*{corollary*}{Corollary}

\usepackage{natbib} 
\bibpunct{(}{)}{;}{a}{}{,} 

\usepackage{etoolbox} 
\apptocmd{\sloppy}{\hbadness 10000\relax}{}{} 

\usepackage{multibib}
\newcites{sec}{References}

\usepackage{color}
\usepackage{listings}
\usepackage{hyperref}
\usepackage{booktabs}

\usepackage{lscape}

\RequirePackage[normalem]{ulem}

\usepackage{bm}

\def \P{\mathbb{P}}
\def \V {\mathbb{V}}
\def \bfE {\mathbb{E}}
\def \E {\mathbb{E}}
 
  \def \cJ{\mathcal{J}}

\def \bA { {\bf A}}
\def \bX {{\bf X}}

\newcommand{\bcol}{\textcolor{black}}

\newcommand{\indep}{\perp \!\!\! \perp}

\allowdisplaybreaks

\begin{document}

\singlespacing

\title{\bf  Matching-Based Nonparametric Estimation of Group Average Treatment Effects}

\author[1]{Peng Wu}
\author[1]{Pengtao Zeng}
\author[1]{Zhaoqing Tian}
\author[2]{Shaojie Wei\thanks{Corresponding author: weishaojie@bwu.edu.cn}}
\affil[1]{\small School of Mathematics and Statistics, Beijing Technology and Business University, 100048, China}
\affil[2]{\small School of Systems Science and Statistics, Beijing Wuzi University, Beijing, 101126, China}


\date{}

\maketitle

\begin{abstract}
Heterogeneous treatment effects, which vary according to individual covariates, are crucial in fields such as personalized medicine and tailored treatment strategies.
In many applications, rather than considering the heterogeneity induced by all covariates, practitioners focus on a few key covariates to develop tailored treatment decisions. Based on this, we aim to estimate the group average treatment effects (GATEs), which represent heterogeneous treatment effects across subpopulations defined by certain key covariates. Previous strategies for estimating GATEs, such as weighting-based and regression-based methods, suffer from instability or extrapolation bias, especially when several propensity scores are close to zero or one. To address these limitations, we propose two novel nonparametric estimation methods: a matching-based method and a bias-corrected matching method for estimating GATEs.
The matching-based method imputes potential outcomes using a matching technique, followed by a nonparametric regression. This method avoids the instability caused by extreme propensity scores but may introduce non-negligible  bias when the dimension of full covariates is high. To mitigate this, the bias-corrected matching estimator incorporates additional outcome regression models, enhancing robustness and reducing bias. We show the consistency, double robustness, and asymptotic normality of the bias-corrected matching estimator. We empirically demonstrate the advantages of the proposed methods with extensive simulation studies and a real-world application. An open-source \textsf{R} package, \texttt{MatchGATE}, is available to implement the proposed methods.
\end{abstract}


\medskip
\noindent
{\bf Keywords}:
Causal Inference, Group Average Treatment Effects, Heterogeneity, Nonparametric

\newpage

\onehalfspacing


\section{Introduction} \label{sec:1}

The variation in treatment effects due to differing characteristics, known as heterogeneous treatment effect, is widely applied in personalized medicine~\citep{Kent-etal2018, Kosorok+Laber:2019}, policy learning~\citep{murphy2003optimal,2021Policy}, and tailored marketing strategies~\citep{Imai-Strauss2011, Yin2018}. In many instances, rather than considering the heterogeneity induced by all covariates, researchers focus on a few core characteristics relevant to treatment to develop tailored treatment decisions~\citep{Chakraborty-Moodie2013, Lee-Okui-Whang-2017, Chernozhukov-Luo2018, Semenova-Chernozhukov}.
For example, in clinical practice, older patients tend to experience more side effects. Consequently, physicians often recommend adjusting drug efficacy based on age to balance the benefits and risks of treatment~\citep{Velentgas-etal2013}.
In health policy, age-dependent vaccine effectiveness has been utilized to guide targeted vaccination programs~\citep{Soiza-etal2021}.
This article is motivated by studies on the treatment effects of biologic therapy for psoriasis.
With the inclusion of biologics therapy in China's national drug reimbursement list, concerns have arisen regarding its appropriate use across different age groups. This necessitates the development of methods to distinguish the heterogeneity of biologics therapy effects across various age groups from the effects induced by the remaining covariates.

For clarity, we formally present the quantity of interest.
 Let $Y$ be the outcome, $A$ be a binary treatment taking values in $\{0, 1\}$, $X$ be the full baseline covariates, and $Z$ be the key covariates  used to define the subgroups of interest.  The covariates $Z$ can be any covariates in $X$, or a given real-valued function of $X$.  Denote $Y^0$ and $Y^1$  as the potential outcomes under treatment arms zero and one, respectively.  In this article, we focus on the group average treatment effect (GATE) $\tau(z)$,
  defined as $\E(Y^{1} - Y^{0} \mid Z=z)$ for possible values $z$.
   Compared to the conditional treatment effects given the full covariates (i.e., $\tau(x) = \E(Y^1 -Y^0\mid X=x)$)~\citep{Nie-Wager-2018, Wager-Athey-2018}, the GATE $\tau(z)$ are more easily interpretable and
   have seem broader usages in clinical settings~\citep{Abrevaya-Hsu-Lieli-2015, Zimmert-Lechner-2019, Wu-Tan2021}.

However, estimating GATEs presents challenges. The methods used must be capable of flexibly distinguishing the heterogeneous effect caused by $Z$, from the effects induced by the remaining covariates. In addition, these remaining covariates can still confound the effect of treatment and outcome, merely conditioning on $Z$ is insufficient to control for all confounding. Propensity score-based weighting methods have been investigated for estimating GATEs.
For instance, Abrevaya et al. \cite{Abrevaya-Hsu-Lieli-2015} introduced the inverse probability weighting (IPW) estimator for GATE, which utilizes the inverse of propensity scores as weights to adjust outcomes. However, the IPW estimator can exhibit high instability when propensity scores approach zero or one, a common issue observed in weighting methods for estimating average treatment effects~\citep{Hahn-1998, Rubin-2001, Kang-Schafer-2007}.
By incorporating additional parametric outcome regression models, Lee et al. \cite{Lee-Okui-Whang-2017} proposed the augmented inverse probability weighting (AIPW) estimator for GATE.
The AIPW estimator is doubly robust, which remains consistent if either the propensity score model is correctly specified or the outcome regression models are correctly specified.
In addition, Fan et al. \cite{Fan-Hsu-Lieli-Zhang-2019}, Zimmert and Lechner \cite{Zimmert-Lechner-2019} and Semenova
and Chernozhukov \cite{Semenova-Chernozhukov} extended the AIPW estimator of Lee et al. \cite{Lee-Okui-Whang-2017} by employing machine learning methods to estimate the propensity score and outcome regression functions.
Nevertheless, the AIPW methods also use the inverse of propensity scores as weights and can be very volatile when propensity score values are extreme (close to zero or one). Additionally, estimating outcome regression functions using machine learning methods heavily relies on extrapolation~\citep{Wu-Han2022}. In cases where the covariate balance between the treated and control groups is poor, outcome regression functions estimated through extrapolation may exhibit significant bias empirically, leading to unreliable conclusions for GATE, see Section \ref{sec2-2-1} for a detailed discussion.

\begin{table}[t]
\centering
\caption{Comparison of various estimators of GATE, where symbols $\checkmark$ and  $\times$ denote yes and no, respectively. The matching and bias-corrected matching are the proposed estimators in this article. A detailed discussion of these estimators is provided in Sections \ref{look} and \ref{sec3}.}
\small
\begin{tabular}{l| c c c c c c}
\toprule
	 &  IPW  &  OR &    PSR  & AIPW  &  Matching   & Bias-Corrected Matching     \\ \midrule
		Doubly robust  & $ \times$ & $ \times$   &     $ \times$  &  {\checkmark}  & $ \times$    & {\checkmark}             \\
		Robust to extreme propensity scores &  $ \times$  & \checkmark  & \checkmark  &   $\times$  & {\checkmark} & \checkmark         \\
		Without model extrapolation &  {\checkmark} &  $\times$ &  \checkmark &  $\times$  &   \checkmark  &    $\times$    \\
\bottomrule
	\end{tabular}  \label{tab1}
\end{table}

Alternative methods for estimating GATEs involve a two-step regression process. These methods first estimate treatment effects at a more granular level, and then aggregate it to estimate the GATE.
For example, the outcome regression (OR) method initially estimates $\tau(X)$ as the treatment effect at a lower level of granularity, and then further regress $\tau(X)$ on $Z$ to estimate $\tau(z)$.
However, since the dimension of $X$ is usually much larger than that of $Z$,  it is typically difficult to non-parametrically estimate the conditional treatment effect based on the full covariates~\citep{Abrevaya-Hsu-Lieli-2015, Zimmert-Lechner-2019, Wu-Tan2021}.
To address this issue, Wu et al. \cite{Wu-Han2022} proposed an alternative estimand for the first-step estimation. This approach aims to find an estimand with a granularity level higher than $\tau(X)$ but still lower than $\tau(Z)$. Compared to inverse probability weighting methods, this new strategy is less sensitive to extreme propensity scores as it incorporates propensity scores as nonparametric regressors. However, for inference, it heavily depends on the model specification of the propensity score, see Section \ref{sec2-2-2} for more details.

In this article, we propose two novel methods for estimating GATEs: a matching-based nonparametric estimator and a bias-corrected nonparametric estimator.
Table \ref{tab1} compares the various GATE estimators in terms of three desirable properties: double robustness, robust to extreme propensity scores (i.e., propensity scores close to zero or one), and  without model extrapolation.
It indicates that both the proposed estimators  complement the methodology for estimating GATEs.
The proposed matching-based nonparametric estimator first imputes the potential outcomes for each unit using the matching technique~\citep{Abadie-Imbens-2006}, and then performs a nonparametric local constant regression of the contrast of the imputed potential outcomes on $Z$ to obtain the estimator of GATE. This estimator is fully nonparametric and does not rely on the inverse of propensity scores as weights, making it robust to extreme propensity scores. However, the matching procedure introduces a non-negligible error depending on the dimension of the covariates $X$. When the dimension of $X$ is 
large, the convergence rate of the proposed matching-based estimator may result in significant bias empirically.
 To deal with this problem, we further propose the bias-corrected matching estimator by introducing an additional outcome regression model. We show the asymptotic properties of the proposed methods, including consistency, double robustness, and asymptotic normality.

The rest of the paper is organized as follows. In Section \ref{sec2}, we describe the formal setup, basic assumptions and a summary of previous methods for estimating GATE. The review of the state-of-the-art estimation methods provides a strong motivation for the paper. In Section \ref{sec3}, we formally propose the matching-based estimator and the bias-corrected matching estimator. A detailed exploration of the theoretical properties of the proposed methods is also given. Section \ref{sec4} presents the finite-sample performance of our methods through simulation studies. Section \ref{sec5} illustrates the proposed methods by an application to real-world data on biologics therapy for psoriasis. Section \ref{sec6} concludes with a discussion. Technical proofs are relegated to the Appendix.

\section{Background} \label{sec2}

\subsection{Setup}

Let $A$ denote the indicator for binary treatment, with $A=1$ or 0 the treated or control group, $X \in \mathcal{X} \subset \mathbb{R}^p$ denote the observed pre-treatment covariates,  $Y \in \mathcal{Y} \subset \mathbb{R}$ denote the observed outcome of interest. Under the potential outcome framework \citep{Rubin1974, Neyman1990},  let $Y^1$ and $Y^0$  be the potential outcomes if the active treatment ($A=1$) and control treatment ($A=0$) are received, respectively. Under stable unit treatment value assumption \cite{Rubin1980}, i.e., there are no multiple versions of treatment and no interference across units, observed outcome $Y$ can be represented as $Y = AY^1 + (1 - A)Y^0$.
		
Let $Z$ be the covariates used to define subgroups of interest, the group average treatment effect (GATE) 
  is defined as
			\[   \tau(z) = \bfE(Y^1 - Y^0 \mid Z=z),    \]
which measures the average treatment effect conditional on the subgroup of $Z=z$. The covariates $Z$ can be any subvector of $X$, or a given real-valued function of $X$.
For example, in our empirical application,
we examine the effects of biologics versus conventional therapies on psoriasis area and severity (PASI) improvement in subpopulations defined by a patient's age.
For identification of $\tau(z)$,  we maintain the following common Assumption \ref{assump1} throughout.
\begin{assumption} \label{assump1}
(i) Unconfoundedness: $A \indep (Y^{0}, Y^1) \mid X$ \citep{Rubin1976}; (ii) Overlap: $0 < \pi(x) < 1$ for all $x \in \mathcal{X}$, where $\pi(x) = \P(A=1 \mid X= x)$ is the propensity score \citep{Rosenbaum-Rubin-1983}.
\end{assumption}

The unconfoundedness assumption is typically adopted in causal inference with observational data, asserting that the potential outcomes and treatment are independent conditioning on the covariates. This ensures that there was no difference in the distributions of potential outcomes between the groups of $A=1$ and the group of $A=0$ conditional on $X$ and implicitly rules out the existence of unmeasured confounders affecting both treatment and outcomes.
The overlap assumption is sometimes referred to as "positivity" because it ensures that all units have a positive probability of receiving each treatment~\citep{Hernan-Robins-2020}. In addition, it guarantees that the support sets of covariates in the treatment group overlap with the support sets of covariates in the control group.
With Assumption \ref{assump1}, we can identify $\tau(z)$ as
		\begin{equation}      \label{eq1}
\tau(z)
= \E\{  \E(  Y^1 - Y^0 \mid X)  \mid Z = z \}
= \E\{  \E(  Y \mid X, A =1)- \E( Y \mid X, A =0)  \mid Z= z\},
\end{equation}
where the first equality follows from the law of iterated expectations, and the second follows by Assumption \ref{assump1}.

From equation (\ref{eq1}), $\tau(z) = \E\{\tau(X) \mid Z=z\}$ with $\tau(X) = \E(Y^1 - Y^0 \mid X)$. This indicates that  $\tau(Z)$ is at a higher level of granularity than the treatment effects conditional on the full covariates.  Observational studies often include a large set of covariates, possibly with nonlinear and interaction terms,  to reduce confounding bias and enhance the credibility of causal inference.  Thus, the dimension of $X$ is typically much higher than that of $Z$.  By conditioning on a low-dimensional covariate,
  $\tau(z)$ is easier to interpret in practice.  Moreover, estimating $\tau(z)$ is more manageable and less affected by modeling assumptions in statistical analysis.

Suppose that the observed data $\{ (Y_i, A_i, X_i): i = 1, ..., N \}$ consists of independent and identically distributed (i.i.d.) sample of $N$ observations from a super population. Let $\mu_a(X) = \E(Y\mid X, A = a)$ for $a = 0, 1$ be the outcome regression functions.  For notational simplicity, we focus on the case where $Z$ is a scalar and the proposed methods can readily extend to the case of multi-dimensional $Z$.

\subsection{A closer look at previous methods} \label{look} 

In recent years, there has been a growing interest in estimating $\tau(z)$. In this subsection, we review several state-of-the-art estimation methods, describing their basic ideas, strengths, and limitations, thereby providing a strong motivation for this work.



\subsubsection{Propensity score based  weighting methods} \label{sec2-2-1}
Several methods use propensity scores to achieve the covariate balance by reweighing the units with the inverse of propensity scores.
 For example,  Abrevaya et al.
 \cite{Abrevaya-Hsu-Lieli-2015}  noted that
 \begin{equation} \label{eq:IPW}
 \tau(z) =  \mathbb{E} \Big [  \frac{AY}{\pi(X)} - \frac{(1-A)Y}{1-\pi(X)}   \Big | Z = z   \Big ]
 \end{equation}
 and based on this the authors obtained the inverse probability weighted (IPW) estimator of $\tau(z)$ as follows:
(a) estimate the propensity score $\pi(X)$, denoted as  
$\hat \pi(X)$;
(b) construct an estimate of $\tau(z)$ by using nonparametric regression
$AY/\hat \pi(X) - (1-D)Y/(1- \hat \pi(X))$ on $Z$. For example, a local constant estimator of $\tau(z)$ is  given as
     \[  \hat \tau(z)^{ipw} =       \sum_{i=1}^{N} K( \frac{Z_{i} - z}{h})\left \{  \frac{A_i Y_i}{\hat \pi(X_i)} - \frac{(1-A_i)Y_i}{1-\hat \pi(X_i)}   \right \}  \Big /  \sum_{i=1}^{N} K( \frac{Z_{i} - z}{h}),     \]
where $h$ is a bandwidth and $K(\cdot)$ is a symmetric kernel function~\citep{Li-Racine-2007, Fan-Gijbels1996} that satisfies $\int K(t)dt = 1$ and $\int t K(t)dt =0$.  For example, Epanechnikov kernel $K(t) = 3(1-t^2)\mathbb{I}\{|t|\leq 1\}/4$ and Gaussian kernel $K(t)= \exp(-t^2/2)/\sqrt{2 \pi}$ for $t\in \mathbb{R}$.
As in weighting methods \cite{Hahn-1998, Rubin-2001, Kang-Schafer-2007}, the IPW estimator based on equation  (\ref{eq:IPW}) is 
 highly unstable if the propensity score values are close to zero or one. Additionally, for inference,  the IPW estimator relies heavily on the model specification of the propensity score.

Subsequently,   Lee et al. \cite{Lee-Okui-Whang-2017} extended the IPW estimator and proposed the augmented IPW (AIPW) estimator of $\tau(z)$ based on the following equality 
 \begin{equation} \label{eq:AIPW}
 \tau(z) =  \E \left [ \frac{A(Y - \mu_1(X))}{\pi(X)} - \frac{(1-A)(Y - \mu_0(X))}{1-\pi(X)} +  ( \mu_1(X) - \mu_0(X) ) \Big | Z = z    \right ].
 \end{equation}
 Let $\hat \pi(X)$ and $\hat \mu_a(X)$ be the estimates of $\pi(X)$ and $\mu_a(X)$ for $a = 0, 1$.
 Then, equation (\ref{eq:AIPW}) implies that the AIPW estimator of $\tau(z)$ can be obtained by regressing  $A(Y - \hat \mu_1(X))/\hat \pi(X) - (1-A)(Y - \hat \mu_0(X))(1-\hat \pi(X)) +  (\hat \mu_1(X) -\hat \mu_0(X) )$ on $Z$, e.g., a local constant estimator is given as
         \begin{align*}
            &\hat \tau(z)^{aipw} \\
             =&       \sum_{i=1}^{N} K( \frac{Z_{i} - z}{h})\left \{  \frac{A(Y -\hat \mu_1(X))}{\hat \pi(X)} - \frac{(1-A)(Y - \hat \mu_0(X))}{1-\hat \pi(X)} +  (\hat \mu_1(X) - \hat \mu_0(X) ) \right \}  \Biggl /  \sum_{i=1}^{N} K( \frac{Z_{i} - z}{h}).
         \end{align*}
         
 The AIPW estimator enjoys the doubly robust property by combining the outcome regression model and propensity score model.   An estimator of $\tau(z)$ is doubly robust in the sense that it is consistent when either the outcome regression model or the propensity model is correctly specified~\citep{Bang-Robins-2005}.

 Compared with the IPW estimator, the AIPW estimator, in addition to being doubly robust, potentially offers a further advantage by allowing flexible machine learning methods to estimate nuisance parameters $(\pi(X), \mu_0(X), \mu_1(X))$.
       Since both $\hat \mu_{a}(\cdot)$ and $\hat \pi(\cdot)$ may converge at a slower rate than $O_{\P}(1/\sqrt{nh})$ if machine learning methods are adopted,
        the resulting convergence rates for the IPW  estimator will be slower than $O_{\P}(1/\sqrt{nh})$.
However, the AIPW estimator could attain $\sqrt{nh}$-consistent only if both models $\pi(X)$ and $\mu_a(X)$ are
  correctly specified or with negligible biases typically having a convergence rate faster than $O_{\P}((nh)^{-1/4})$.
  Based on this observation,  several works have tried machine learning methods to avoid model specification for nuisance parameters, and used sample splitting (or cross-fitting) technique to reduce the impact of nuisance parameters estimation on the estimator of $\tau(z)$ \cite{Fan-Hsu-Lieli-Zhang-2019, Semenova-Chernozhukov, Zimmert-Lechner-2019}. In addition, Wu et al. \cite{Wu-Tan2021} extended these works by enhancing the estimation of confidence intervals.

It is expected that both existing IPW and augmented AIPW estimators of $\tau(w)$ 
are very volatile when there are propensity score values very close to zero or one. This is
because the resulting weights (inverse of propensity scores) are extremely large and they tend to inflate the variance \cite{Rubin-2001,Kang-Schafer-2007,Tan2007,Molenberghs-etal2015,Vermeulen-Vansteelandt2015}.
In addition, estimating  $\mu_a(X)$ using machine learning methods heavily on  extrapolation \cite{Kang-Schafer-2007, Tan2007}.  This is because the outcome regression model $\mu_a(X)$ is trained based on the data of $A = a$, while using the predicted values in the whole data.  When the balance of covariates between the treated group and control group is poor, the estimated outcome regression functions based on extrapolation may have a large bias empirically and hence lead to unreliable conclusions. 

\subsubsection{Two-step regression methods} \label{sec2-2-2}

To estimate an estimand at a higher level of granularity, an intuitive way is to estimate the treatment effects at a lower level of granularity first, and then integrate the obtained estimates into the subspace of interest.
 Below, we describe this idea further.
Based on the observation that
   \begin{equation*} \tau(z) = \E[ \tau(X) \mid Z=z ] = \E[  \mu_1(X) -  \mu_0(X) \mid Z=z ], \end{equation*}
  $\tau(X)$ is a natural choice for such an estimand at a lower level of granularity.   Thus, an outcome regression (OR) based local constant estimator of $\tau(z)$ is given as
       \[  \hat \tau(z)^{or} =       \sum_{i=1}^{N} K( \frac{Z_{i} - z}{h})\left ( \hat \mu_1(X_i) - \hat \mu_0(X_i)  \right )  \Big /  \sum_{i=1}^{N} K( \frac{Z_{i} - z}{h}).     \]
 That is,  we can estimate the insider regression function $\mu_1(X) -  \mu_0(X)$ first and then estimate $\tau(z)$ by conducting local constant regression.
In a similar spirit, Lechner \cite{Lechner-2019} proposed a method to construct a modified causal forests estimator of $\tau(X)$,  but when estimating $\tau(z)$, the author averaged $\tau(X)$ across the sample with $Z = z$ directly to obtain the estimator of $\tau(z)$. However, their article focuses on estimating $\tau(x)$, not $\tau(z)$, and the method of taking the average across the sample $Z=z$ directly is only applicable to the case of discrete $Z$.
Similar to the IPW estimator, the OR estimator depends heavily on the model specification of $\mu_a(X)$ for $a = 0, 1$. If flexible machine learning approaches are used to estimate $\mu_a(X)$, the convergence rate of $\hat \tau(z)^{or}$ may be dominated by the convergence rate of $\hat \mu_a(X)$, leading to difficulties in inference. Additionally, like the AIPW estimator, the OR estimator relies heavily on model extrapolation.

An alternative estimand for the first-step estimation was explored by  Wu et al. \cite{Wu-Han2022}  Specifically, they aim to find an estimand that lies in a much higher level of granularity than $\tau(X)$ while still lying in a lower level than $\tau(Z)$, so that in practice it is possible to estimate the new estimand nonparametrically.
The authors choose $\tau(Z, \pi(X)) = \E[ Y^1 - Y^0 \mid Z, \pi(X) ]$ as the estimand, since
 it satisfies
	 \begin{equation*}
	 		\tau(z) = \E (  \tau(Z, \pi(X)) \mid Z = z  ).
				 \end{equation*}
Then the propensity score regression (PSR) method \cite{Wu-Han2022}  consists of three steps: (a) estimate the propensity score with a parametric model; (b) estimate $\tau(Z, \pi(X))= \E[ Y| A=1, Z, \pi(X) ] - \E[ Y| A=0, Z, \pi(X) ]$ by nonparametrically regressing $Y$ on the covariate $Z$ and the estimated propensity score $\hat \pi(X)$; (c) estimate $\tau(z)$ by nonparametrically regressing the estimated $\tau(Z, \pi(X))$ on $Z$.  	
By incorporating propensity scores as nonparametric regressors, the PSR method is less sensitive to extreme propensity scores. However, it depends on the parametric model specification of the propensity score to mitigate the influence of propensity score estimation on the resulting estimator of $\tau(z)$. 

\subsubsection{Motivation: summary of previous methods}

We introduce four desirable statistical properties for evaluating the previous methods comprehensively.
\begin{itemize}
\item \emph{Doubly robust}. The AIPW estimator is doubly robust while the IPW, OR and PSR estimators are not.

\item \emph{Robust to extreme propensity scores.} The propensity score $\pi(X)$ appears in the denominator of both the IPW and AIPW estimators. In the presence of extreme propensity scores, the inverse propensity scores become extremely large and cause much instability. In contrast, the OR and PSR estimators are less prone to this issue.

    \item \emph{Without model extrapolation.} The AIPW and OR estimators rely on model extrapolations of outcome regression functions. In comparison, the estimation of propensity score doesn't rely on model extrapolation.

\end{itemize}

Table \ref{tab1} summarizes the properties of the proposed estimators, where matching and bias-corrected matching are our proposed estimators in this article. The proposed estimators complement the methodology for estimating GATEs.

\subsection{Caveate: propensity score matching fails in estimating GATE} \label{caveate}

Before presenting the proposed method, it's noteworthy that the commonly used propensity score matching method~\citep{Stuart-2010, Abadie-Imbens-2016} fails to yield a consistent estimator for $\tau(z)$. Specifically, we will demonstrate that
\begin{align} \label{eq6}
\tau(z)
\neq{}& \E \left [ \E(Y \mid  \pi(X), A =1 )  - \E(Y \mid \pi(X), A  = 0)  ~ \big | ~   Z = z \right ] = \E \left [ \E(Y^1 - Y^0 \mid  \pi(X))    ~ \big | ~   Z = z \right ].
\end{align}
\begin{proof}
Since $\pi(X)$ is a balancing score , $A  \indep X \mid \pi(X)$. Combine it with the unconfoundedness assumption $A \indep (Y^0, Y^1) \mid \pi(X)$, we have $A  \indep (Y^0, Y^1) \mid   \pi(X), Z$
by the corollary 15.2.1 of Anderson \cite{Anderson-2003}. Then,  
\begin{align*}
\tau(z)
= &\E(Y^1 - Y^0 \mid Z = z )  \\
= &\E[ \E( Y^1 - Y^0 \mid Z, \pi(X))  \mid Z = z ]  \\
= &\E[  \E(Y^1 \mid Z, \pi(X), A = 1) - \E( Y^0 \mid Z,  \pi(X), A = 0)\mid Z = z ] \\
= &\E[ \E(Y \mid Z, \pi(X), A = 1) - \E( Y \mid Z, \pi(X), A = 0) \mid Z = z ].
\end{align*}
Note that
	\[    \E(Y \mid Z, \pi(X), A = 1) - \E( Y \mid Z, \pi(X), A = 0) \neq  \E(Y \mid  \pi(X), A =1 )  - \E(Y \mid \pi(X), A  = 0) \]
in general, which implies equation \eqref{eq6}.
\end{proof}

Essentially,  propensity score matching can only obtain an estimate of GATE at the $\pi(X)$ granularity level, which differs in the GATE at the $Z$ granularity level.

\section{Proposed method}  \label{sec3}
In this section, we propose two nonparametric estimators for the GATE: a matching-based estimator and a bias-corrected matching estimator. Their strengths are summarized in Table \ref{tab1}.

\subsection{Matching-based nonparametric estimator} \label{sec3-1}

\subsubsection{Method}
We introduce a matching-based nonparametric  method to estimate the GATE, utilizing matching ``with replacement',  wherein each unit can be used as a match more than once~\citep{Abadie-Imbens-2006}. Matching with replacement
has two appealing features: (a) it generates higher-quality matches compared to matching without replacement by expanding the pool of potential matches; (b) it performs matching for each unit, preventing any data from being discarded post-matching. This ensures that the post-matching data retains the same covariate distribution as the observed data.

\smallskip
The proposed matching-based method is described as follows.

{\bf Step 1 (Matching).}  Let $||\cdot||$ be a pre-specified distance metric, and $l_{m}(i)$ be the index of the $m$-th match to unit $i$, i.e., $A_{l_{m}(i)} = 1 - A_{i}$ and $\sum_{j: A_{j} = 1- A_{i}} \mathbb{I}\{ || X_{j} -  X_{i}|| \leq ||  X_{ l_{m}(i)} - X_{i}|| \} = m$, where $\mathbb{I}(\cdot)$ is the indicator function.
Let $\cJ_{M}(i) = \{l_{1}(i), ....,  l_{M}(i)\}$ denote the set of indices for the first $M$ matches for unit $i$, then we impute the potential outcomes by averaging the outcomes of $M$ matches, i.e.,
\begin{align*}  \hat Y_{i}^0 ={}& \begin{cases}
		&Y_{i}, \quad \qquad \qquad  \text{if } A_{i} = 0,\\
		& \frac 1 M \sum_{j \in \cJ_{M}(i)} Y_{j},  ~ \text{if } A_{i} = 1,
\end{cases}
\quad
 \hat Y_{i}^1 = \begin{cases}
		& \frac 1 M \sum_{j \in \cJ_{M}(i)} Y_{j},  \text{if } ~ A_{i} = 0,\\
		& Y_{i}  \quad \qquad \qquad ~ \text{if }~ A_{i} = 1.
\end{cases}
   \end{align*}
 The imputation method is intuitive: if $A_i =a$, the imputed potential outcome $\hat Y_i^a$ is equal to the observed outcome $Y_i$;  if $A_i = 1-a$, then $Y_i^a$ is missing, and we impute it using the average of outcomes for the $M$ units that are most similar to the unit $i$ and belonging to group $A=a$. 
     After matching, we obtain the pseudo individual treatment effect $\hat Y_i^1 - \hat Y_i^0$ for each unit $i$.

\smallskip
{\bf Step 2 (Estimation of GATE).}  To avoid any parametric model specification, we propose using nonparametric local constant regression to obtain the estimator of $\tau(z)$, which is given by 
 	\begin{equation}  \label{eq-7}
	     \hat \tau(z)^{match}  =    \sum_{i=1}^{N} K( \frac{Z_{i} - z}{h}) \{ \hat Y_{i}^1- \hat Y_{i}^0  \} \Big /     \sum_{i=1}^{N} K( \frac{Z_{i} - z}{h})
	\end{equation}  	
for continuous $Z$. 	
In addition, when $Z$ is a discrete variable, this reduces to calculating the conditional sample averages stratified by $Z$, that is, replacing $K(\frac{Z_i - z}{h})$ with the indicator function $\mathbb{I}(Z_i = z)$ in equation (\ref{eq-7}).


\medskip

As shown in Table \ref{tab1}, the proposed matching-based estimator has two desirable properties: robust to extreme propensities and without model extrapolation.
This is attributed to its avoidance of using the inverse of propensity score as weight, making it robust to extreme propensities. Moreover, being fully nonparametric eliminates the need for model extrapolation and parametric model specification.

The matching-based estimator is analogous to the OR estimator. The former can be interpreted as imputing $Y_{i}^0$ and $Y_{i}^1$ with nearest-neighbor estimates, while the latter can be interpreted as imputing $Y_{i}^0$ and $Y_{i}^1$ with $\hat \mu_0(X_i)$ and $\hat \mu_1(X_i)$. Nevertheless, compared with the OR estimator, the proposed matching-based estimator does not rely on model extrapolation and parametric model specification. Interestingly, in the next subsection, we will see that this improvement comes at the cost of a slower convergence rate when the dimension of covariates is high.

\subsubsection{Theoretical analysis}

We theoretically analyze the proposed estimator $\hat \tau(z)^{match}$. First, we show its convergence rate, which relies on regularity Assumptions \ref{assum_2}-\ref{assum_3} below.
Let $\sigma_a( x) = \V( Y^a \mid  X=  x )$ and  $\epsilon_{i} = Y_{i} - \mu_{A_i}( X_i)$ for $x \in \mathcal{X}, a \in \{0, 1\}$.

\begin{assumption}[regularity conditions for matching \cite{Abadie-Imbens-2006}] \label{assum_2}
For $a = 0, 1$,  \par
    \begin{itemize}
        \item[] (i)    $\mu_a( x)$ and $\sigma^2_a(  x)$ are Lipschitz continuous of $x \in \mathcal{X}$; $ \mathcal{X}$  is a compact set; 
        \item[] (ii)
        $\E[ (Y)^4 |  X=  x, A = a]$ exist and is bounded uniformly in $x \in \mathcal{X}$;
        \item[] (iii)   $\sigma^2_a( x)$ is bounded away from zero.
    \end{itemize}
\end{assumption}	

\begin{assumption}[regularity conditions for local constant regression, Section 2 of Li and Racine \cite{Li-Racine-2007}]  \label{assum_3}
For any given $z$,
\begin{itemize}
    \item[]  (i)  $f(z)$ is twice differentiable and $f(z) > 0$, where $f(z)$ is the density function of $Z$.
    \item[] (ii) $h \to 0$, $nh \to \infty$ as $n \to \infty$, and $\tau(z)$ is twice differentiable. 
    \item[] (iii) $\E[\epsilon^2| Z =z]$ and $\text{Var}(\tau( X) | Z =z)$ is finite and differentiable.
    \item[] (iv) $K(\cdot)$ is a symmetric kernel function.
\end{itemize}
\end{assumption}

Assumption \ref{assum_2} is the standard assumption to ensure the validity of the matching estimator~\citep{Abadie-Imbens-2006}. Assumption \ref{assum_3} comprises the regular conditions widely employed in local constant regression~\citep{Fan-Gijbels1996, Hardle-2004, Li-Racine-2007}.

\begin{theorem}  \label{thm1} Suppose that $M$ is a fixed constant and Assumptions 1--3 hold, then for any given $z$,  
  \[ \hat \tau(z)^{match} - \tau(z) = O_{\P}(N^{-1/p} + h^2 + 1/\sqrt{Nh} ). \]
\end{theorem}

Theorem \ref{thm1} shows the proposed estimator $\hat \tau(z)$ has a convergence rate of order $ O_{\P}(N^{-1/p} + h^2 + 1/\sqrt{Nh} )$ for a fixed $M$, where the term $O_{\P}(N^{-1/p})$ is induced by matching and the term $O_{\P}(h^2 + 1/\sqrt{Nh})$ is caused by nonparametric local constant regression. In particular,  as $N$ goes to infinity, $\hat \tau(z)$ will converge to $\tau(z)$ in probability, i.e., $\hat \tau(z)$ is a consistent estimator of $\tau(z)$.
Next, we establish the connection between the proposed matching-based estimator $ \hat \tau(z)^{match} $ and the IPW estimator $ \hat \tau(z)^{ipw}$ by allowing $M$ to depend on $N$,
which is based on the following Proposition \ref{prop1}. Let $f(X=x \mid A=0)$ and $f(X=x \mid A=1)$ be the probability density functions of $X$ given $A=0$ and $A=1$, respectively.

\begin{proposition}[Connection to the IPW Estimator] \label{prop1}  We have that

(a)  Under the conditions in Theorem \ref{thm1},
       \begin{align*}
         \hat \tau(z)^{match} ={}&   \sum_{i=1}^{N} K( \frac{Z_{i} - z}{h}) \left \{  A_i Y_i \cdot \left (1+ \dfrac{K_M(i)}{M} \right )  - (1-A_i)Y_i \cdot\left (1+ \dfrac{K_M(i)}{M} \right )    \right \}  \Big /    \sum_{i=1}^{N} K( \frac{Z_{i} - z}{h}) \\ & +   O_{\P}(  N^{-1/p} ),  \end{align*}
     where  $K_M(i)$ is the number of times unit $i$ is used as a match, i.e.,  $K_M(i) = \sum_{l=1}^N I\{i \in \cJ_M(l)\}.$

(b) If $f(X=x \mid A=0)$ and $f(X=x \mid A=1)$ satisfy the Assumption 4.3 of Lin et al. \cite{Lin-etal2023} and $M = c N^{2/(2+p)}$ for some constant $c > 0$, then
       \[              \hat \tau(z)^{match}  =  \sum_{i=1}^{N} K( \frac{Z_{i} - z}{h}) \left (  \frac{A_i Y_i}{ \pi(X_i)} - \frac{(1-A_i)Y_i}{1- \pi(X_i)}   \right ) \Big /    \sum_{i=1}^{N} K( \frac{Z_{i} - z}{h})   +    O_{\P}(  N^{-1/(2+p)} ).  \]
\end{proposition}

Proposition \ref{prop1}(a) provides an equivalent expression for $ \hat \tau(z)^{match} $ using the notation $K_M(i)$, which is the number of times unit $i$ is used as a match. This reformulation aids in analyzing the convergence rate of the matching-based estimator~\citep{Abadie-Imbens-2006, Abadie2011}.  Proposition \ref{prop1}(b) establishes a connection between the matching-based estimator with the IPW estimator when $M$ is allowed to depend on $N$.

We provide a heuristic interpretation for Proposition \ref{prop1}(b). Concretely,
if $M$ relies on $N$, then in accordance with the findings of Lin et al. \cite{Lin-etal2023} (Section 5), we have
  \begin{align}  \label{eq9}
  \begin{cases}
            1+ K_M(i)/M  ={}  1/ \pi(X_i) +  O_{\P}(  N^{-1/(2+p)} ), \quad \qquad \text{if $A_i = 1$} \\
            1+ K_M(i)/M  ={}  1/(1- \pi(X_i)) +  O_{\P}(  N^{-1/(2+p)} ), \quad \text{if $A_i = 0$}.
     \end{cases}
     \end{align}
   Then Proposition \ref{prop1}(b) follows immediately by combining (\ref{eq9}) with  Proposition \ref{prop1}(a).

  It is helpful to compare the matching-based estimator with the IPW estimator. The former achieves covariate balance through direct matching, while the IPW estimator uses the inverse of propensity scores as weights to balance covariate distributions between treated and control units. However, a key practical challenge of propensity score-based weighting methods lies in the estimation of the true propensity, which is rarely known and needs to be estimated from the observed data. Commonly used propensity score estimation methods, such as recursive partitioning tree, artificial neural network, random forest, and boosting~\citep{Setoguchi-2008, Lee-2010}, ignore the balancing property of propensity scores, leading to potential estimation errors in ensuring covariate balance \cite{Imai2014, Anna2022}. Moreover, the IPW estimator exhibits instability in the presence of extreme propensity scores.
  In contrast, the matching-based estimator is not susceptible to such issues.


Regrettably, despite its full nonparametric nature and robustness to extreme propensity scores, the matching procedure employed in the proposed matching-based method introduces a non-negligible error of order $O_{\P}(N^{-1/p})$ or $O_{\P}(N^{-1/(2+p)})$. When $p$ is large, 
the convergence rate of $\hat \tau(z)$ becomes dominated by the estimation error induced by the matching process. To address this issue, in Section \ref{sec3-2}, we will introduce an enhanced method that combines the strengths of the proposed matching-based approach with the outcome regression method.

\subsection{Bias-corrected matching-based nonparametric estimator}  \label{sec3-2}

To accelerate the convergence rate of the matching-based estimator $\hat \tau(z)^{match}$, we further propose a bias-corrected nonparametric estimator by introducing additional outcome regression models.
	
\subsubsection{Method}

The motivation for the proposed bias-corrected matching estimator is drawn from Abadie and Imbens  \cite{Abadie2011} and Lin et al. \cite{Lin-etal2023}, and it involves the following three steps.

\smallskip
{\bf Step 1 (Matching).} The same as the matching-based method described in Section \ref{sec3-1}.

{\bf Step 2 (Adjustment of Matching with Outcome Regression Models).}   Let $\hat \mu_{a}(x)$ be a consistent estimator of $\mu_{a}(x)$. Define
\begin{align*} \tilde Y_{i}^0 ={}& \begin{cases}
		&Y_{i}, \quad \qquad \qquad \qquad \qquad \qquad \qquad   \quad\text{if } A_i = 0,\\
		& \frac 1 M \sum_{j \in \cJ_{M}(i)} \{ Y_{j}  + \hat \mu_{0}(X_{i}) - \hat \mu_{0}(X_{j})  \}, \quad  ~~ \text{if } A_i = 1,
\end{cases}   \\
 \tilde Y_{i}^1 ={}& \begin{cases}
		& \frac 1 M \sum_{j \in \cJ_{M}(i)} \{ Y_{j} + \hat \mu_{1}(X_{i}) - \hat \mu_{1}(X_{j})  \}, \quad  ~~  \text{if } A_i = 0,\\
		& Y_{i}  \quad \qquad \qquad \qquad \qquad \qquad \qquad   \quad  ~~ \text{if } A_i = 1.
\end{cases}
   \end{align*}
Here, we require that $\hat \mu_a(x)$ is estimated via sample splitting (or cross-fitting) technique~\citep{Chernozhukov-etal-2018, Wager-Athey-2018}.

{\bf Step 3 (Estimation of GATE).}  Based on the adjusted imputed potential outcomes $\tilde Y_{i}^0$ and $ \tilde Y_{i}^1$, the bias-corrected matching-based (MATCH.bc) local  constant regression estimator is given as
 	\begin{equation*}
	     \hat \tau(z)^{bc}  =  \sum_{i=1}^{N} K( \frac{Z_{i} - z}{h}) \{ \tilde Y_{i}^1- \tilde Y_{i}^0  \} \Big /  \sum_{i=1}^{N} K( \frac{Z_{i} - z}{h}),
	\end{equation*}   	


\medskip
The bias-corrected estimator has an additional adjustment step (Step 2) compared to the matching-based estimator. The rationale behind it is as follows.  When $A_i = 0$,
			\begin{align}
				\tilde Y_{i}^1 ={}&  \frac 1 M \sum_{j \in \cJ_{M}(i)} Y_{j}^1  + \Big \{ \hat \mu_{1}(X_{i}) - \frac 1 M \sum_{j \in \cJ_{M}(i)}\hat \mu_{1}(X_{j})  \Big \} \notag \\
				={}&  \hat Y_{i}^1 +   \Big \{ \hat \mu_{1}(X_{i}) - \frac 1 M \sum_{j \in \cJ_{M}(i)}\hat \mu_{1}(X_{j})  \Big \}     \label{eq11}  \\
				={}& \hat \mu_1(X_i)  +  \Big \{ \frac 1 M \sum_{j \in \cJ_{M}(i)} (Y_{j}^1 - \hat \mu_{1}(X_{j}) ) \Big \},   \label{eq12}
			\end{align}
from which we have three observations:
 \begin{itemize}
  \item Equation (\ref{eq11})  implies that $\tilde Y_{i}^1$ (bias-corrected) adjusts $\hat Y_{i}^1$ (matching) using the outcome regression function $\mu_1(X)$.  Let  $ \bX$ be the matrix with $i$-th row equal to $ X_i^\top$. Note that $\E[\hat Y_i^1 | \bX ] = \frac 1 M \sum_{j \in \cJ_{M}(i)} \mu_{1}(X_{j})$, if the estimated regression function is a good approximation to the true regression function, then $\tilde Y_{i}^1$ will improve $\hat Y_{i}^1$ since the former is closer to $\mu_1(X_i)$.

  \item Equation (\ref{eq12})  implies that $\tilde Y_{i}^1$  adjusts $\hat \mu_1(X_i)$ (outcome regression) using the matching estimate $\frac 1 M \sum_{j \in \cJ_{M}(i)} Y_{j}^1 $.

  \item  Importantly,  even if the outcome regression model is misspecified, i.e., $\hat \mu_1(X_i)$  is not a consistent estimator of $\hat \mu_1(X_i)$. Due to matching, the distance $||X_i - X_j||$ for $j \in \cJ_M(i)$ will converge to zero in probability, leading to $\hat \mu_{1}(X_{i}) - \frac 1 M \sum_{j \in \cJ_{M}(i)}\hat \mu_{1}(X_{j})$ typically being small. Consequently, the resulting bias-adjusted estimator is also consistent under mild conditions.

 \end{itemize}
We can perform a similar analysis of $\tilde Y_i^0$ when $A_i = 1$.
 In summary,  the bias-corrected estimator improves the matching-based estimator if the outcome regression model is correctly specified.  However, if the outcome regression model is misspecified, the bias-corrected estimator behaves similarly to the matching-based estimator. \bcol{Regarding the choice of distance metric in the proposed matching step, we note that while no single metric may be universally optimal, our  matching-based  estimators exhibit only minor performance differences across a variety of commonly used distance metrics (please see Simulation (Section 4), where the performance of the proposed estimators is evaluated under different distance metrics). This highlights the practical flexibility of our methods with respect to metric choice. In practice, one can select Euclidean distance as the distance metric.}

\subsubsection{Theoretical analysis}

In this subsection, we show the large-sample properties of the bias-corrected nonparametric estimator $\hat \tau(z)^{bc}$,
including double robustness, convergence rate, and asymptotic normality.

\smallskip
First,  we show the double robustness of  $\hat \tau(z)^{bc}$.

\begin{proposition}[Double Robustness]   \label{prop2}  Under Assumptions 1--3, suppose that $f(X=x\mid A=0)$ and $f(X=x\mid A=1)$ satisfy the Assumption 4.3 of Lin et al.\cite{Lin-etal2023}, 
  then $\hat \tau(z)^{bc}$ is an consistent estimator of $\tau(z)$ if one of the following conditions holds:

   (i)  $M (\log N) / N \to 0$ and $M \to \infty$ as $N \to \infty$,

    (ii) $||\hat \mu_a - \mu_a ||_{\infty} = o_{\P}(1)$ for $a = 0, 1$,  where  $||\cdot||_{\infty}$ denotes the function $L_{\infty}$ norm.
\end{proposition}

The condition \ref{prop2}(i) ensures that
            $1+ K_M(i)/M  =  1/\pi(X_i) +  o_{\P}(1) \text{ if $A_i = 1$}$, and
            $1+ K_M(i)/M =  1/(1- \pi(X_i)) +  o_{\P}(1) \text{ if $A_i = 0$}$.
  This implicitly indicates the consistency of the nearest neighbor matching-based estimator, given the consistency of the IPW estimator and Proposition \ref{prop1}.
  The condition \ref{prop2}(ii)
   assumes the consistency of the outcome regression models.

Second, we establish the connection between the bias-corrected estimator and the  AIPW estimator, which is helpful to
 obtain the convergence rate and asymptotic normality of $\hat \tau(z)^{bc}$.

\begin{proposition}[Connection to the AIPW Estimator] \label{prop3}
If $f(X=x\mid A=0)$ and $f(X=x\mid A=1)$ satisfy the Assumption 4.3 of Lin et al. \cite{Lin-etal2023}, and $M = c N^{2/(2+p)}$ for some constant $c > 0$, then
           \begin{align*}      &\hat \tau(z)^{bc}  \\={}& \sum_{i=1}^{N} K( \frac{Z_{i} - z}{h}) \Big \{  \dfrac{A_i (Y_i - \hat \mu_1(X_i))}{ \pi(X_i)} - \dfrac{(1-A_i)(Y_i - \hat \mu_0(X_i))}{1- \pi(X_i)} +  (\hat \mu_1(X_i) - \hat \mu_0(X_i) ) \Big \} \Big /    \sum_{i=1}^{N} K( \frac{Z_{i} - z}{h})      \\
            {}&  +    O_{\P}\left (  N^{-1/(2+p)} \cdot ( || \hat \mu_0 - \mu_0  ||_{\infty} +  || \hat \mu_1 - \mu_1  ||_{\infty} ) \right ).
             \end{align*}
\end{proposition}

Based on Proposition \ref{prop3}, we have the following conclusions.

\begin{theorem}[Convergence Rate and Asymptotic Normality] \label{thm2}  Under the conditions of  Proposition \ref{prop3}, we have that

(a) $\hat \tau(z)^{bc}  - \tau(z) = O_{\P}(N^{-1/(2+p)} \cdot ( || \hat \mu_0 - \mu_0  ||_{\infty} +  || \hat \mu_1 - \mu_1  ||_{\infty} ) + h^2 + 1/\sqrt{Nh} )$.

(b) If further  have $|| \hat \mu_a - \mu_a ||_{\infty} = o_{\P}(  N^{-p/(4+2p)} h^{-1/2} )$ for $a = 0, 1$,  then
     \begin{align*}
      &\hat \tau(z)^{bc}\\  =&  \sum_{i=1}^{N} K( \frac{Z_{i} - z}{h}) \Big \{  \dfrac{A_i (Y_i - \mu_1(X_i))}{ \pi(X_i)} - \dfrac{(1-A_i)(Y_i -  \mu_0(X_i))}{1- \pi(X_i)} +  ( \mu_1(X_i) - \mu_0(X_i) ) \Big \} \Big /    \sum_{i=1}^{N} K( \frac{Z_{i} - z}{h}) \\ &+ o_{\P}(1/\sqrt{Nh}),
     \end{align*}
     
   and
   \[  \sqrt{Nh} \cdot \{ \hat \tau(z)^{bc}  -\tau(z) - \text{bias}( \hat \tau(z)^{bc} ) \} \xrightarrow{d} N(0, \sigma^2), \]
  where $\sigma^2 =  \int K(u)^2 du \cdot  \E[  (\frac{A (Y - \mu_1(X))}{ \pi(X)} - \frac{(1-A)(Y -  \mu_0(X))}{1- \pi(X)} +  ( \mu_1(X) -  \mu_0(X) )   - \tau(z) )^2 \mid Z=z ] \big / f(z)$ and  $\text{bias}( \hat \tau(z)^{bc} ) = O(h^2)$ is a bias term induced by local constant regression. 
\end{theorem}



The condition $|| \hat \mu_a - \mu_a ||_{\infty} = o_{\P}(  N^{-p/(4+2p)} h^{-1/2} )$ in Theorem \ref{thm2}(b) only restricts the convergence rate of the estimated outcome regression functions. Typically, for a scalar $Z$, the optimal bandwidth $h = O(N^{-1/5})$. In such a case, $N^{-p/(4+2p)} h^{-1/2}$ equals to $N^{-1/15}$, $N^{-3/20}$, and $N^{-19/60}$ for $p = 1, 2,$ and 10, respectively. Thus, the  condition $|| \hat \mu_a - \mu_a ||_{\infty} = o_{\P}(  N^{-p/(4+2p)} h^{-1/2} )$ is satisfied if  $\hat \mu_a$ has a convergence rate faster than $N^{-1/3}$ for $p \leq 10$.
In addition, similar to the OR and matching-based estimators, the bias-corrected estimator is robust to extreme propensity scores, see Table \ref{tab1} for a summary of the properties of the bias-corrected estimator.

For obtaining the confidence intervals of $\hat \tau(z)^{match}$ and $\hat \tau(z)^{bc}$,  we adopt the subsampling method to estimate the confidence intervals. To do so, we randomly sampled $L$ samples from the full data with a sample size of $N_0^r$ control and $N_1^r$ treated units, where $N_0$ and $N_1$ represent the initial size of control and treated individuals, and $r = 2/3$ as suggested in the literature~\citep{Abadie-Imbens-2008}.  We repeated the estimation method $B$ times for each of the $L$ samples. Variances and confidence intervals of the effects were then calculated as the empirical variances and confidence intervals over $B$ estimates.

\section{Simulation}\label{sec4}
In this section, simulation studies are conducted to evaluate the finite-sample performance of the proposed methods. Specifically, we compare the proposed methods with several potentially competing methods, including the IPW, OR, AIPW and PSR methods, and our proposed matching-based methods are denoted as MATCH and MATCH.bc.


\subsection{Simulation settings}

\begin{figure}
    \centering
    \includegraphics[width=0.9\linewidth]{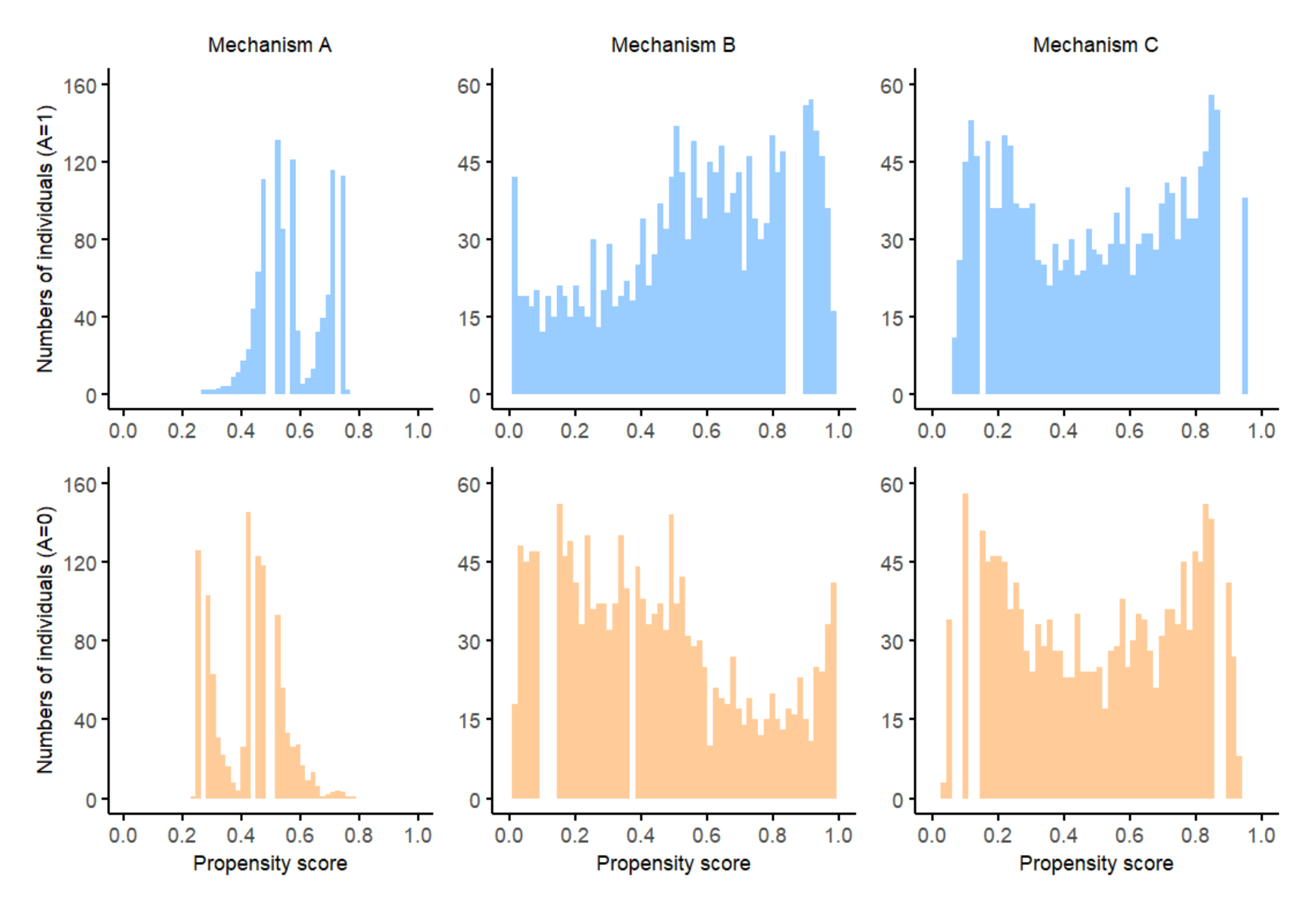}
    \caption{Distributions of propensity score under the three treatment assignment mechanisms ($n = 2000, p = 3$)}
    \label{fig-PS}
\end{figure}

We consider a simple setting of three covariates $X = (X_1, X_2, X_3)$, with $X_1 \sim \text{Unif}(-1/2, 1/2)$, $X_2$ takes values of $\{0, 1, 2\}$ with equal probabilities, and $X_3 \sim \text{Norm}(0, 1)$.  The involved three mechanisms for generating the treatment are set as follows:
    \begin{itemize}
        \item[] {\bf Mechanism A:} $\P(A=1|{X}) = \{1 + \exp( - ((X_1)^2, (X_2)^2, (X_3)^2){\pmb \alpha} ) \}^{-1}$, with ${\pmb \alpha} = (1/2, 1/4, -1/8)^{\rm T}$.
        \item[] {\bf Mechanism B:} $\P(A=1|{X}) = \{1 + \exp( - ((X_1)^2, (X_2)^2, (X_3)^2){\pmb \alpha} ) \}^{-1}$, with ${\pmb \alpha} = (8, 1/2, -5/4)^{\rm T}$.
        \item[]  {\bf Mechanism C:} $\P(A=1|{X}) = \{1 + \exp( - (X_1, X_2, X_3){\pmb \alpha} ) \}^{-1}$, with ${\pmb \alpha} = (5, 1/4, -1/8)^{\rm T}$.
    \end{itemize}
\bcol{Figure \ref{fig-PS} presents histograms of the true propensity scores under the three mechanisms. These histograms are based on a simulation with a sample size of 2,000, illustrating the distributions of the true propensity scores.}
We observe that the propensity scores are not close to 0 or 1 under Mechanism A, and there are some propensity scores close to 0 or 1 under Mechanisms B and C.
The comparison of simulation results under Mechanisms A and B aims to evaluate the estimation performance of different methods when extreme values occur in the propensity score. Mechanisms B and C are established to consider the impacts of the propensity score model specification. The former includes higher-order terms of covariates, while the latter does not.

Let $g(X) = X_2 + X_1X_2 + \frac{1}{2}((X_3)^3 + X_3)$, we considered different studies for generating potential outcomes as follows:
    \begin{itemize}
        \item[] {\bf Study I (Quadratic):} $Y (1) = g({X}) +  2 (X_1)^2 + \epsilon_1$, $Y (0) = g({X}) +  \epsilon_0$.

        \item[] {\bf Study II (Polynomial):}  $Y (1) = g({X}) + X_1(1+2X_1)^2(X_1-1)^2 + \epsilon_1$, $Y (0) = g({X}) +  \epsilon_0$.

        \item[] {\bf Study III (Complex):} $Y (1) = g({X}) + \cos(3 X_1) \log(X_1+2) \exp(X_1) + \epsilon_1$, $Y (0) = g({X}) +  \epsilon_0$.
    \end{itemize}
We chose $Z=X_1$ to define the group of interest, and then, the group average treatment effects $\tau(Z)$ under Studies I, II, and III are respectively represented by quadratic, polynomial, and complex functions of $Z$. Specifically, they are expressed as $2 X_1^2$, $X_1(1+2X_1)^2(X_1-1)^2$, and $\cos(3X_1) \log(X_1 + 2) \exp(X_1)$, respectively. It is worth noting that the true group average treatment effects all have complex forms under Studies I, II, and III, making it challenging to model them correctly. By combining the data generation mechanisms of the treatment and the potential outcomes in pairs, we consider nine possible cases (C1)-(C9) in this section.
The nine cases are classified as follows: in cases (C1)-(C3), Studies I-III use the assignment Mechanism A; in cases (C4)-(C6), Studies I-III use the assignment Mechanism B; in cases (C7)-(C9), Studies I-III use the assignment Mechanism C.


Given the complexity of the true data generation models, researchers are unlikely to correctly specify the true model commonly when estimating GATE in practice. Thus, we adopt common strategies from real-world scenarios during fitting. More concretely, the treatment model is fitted as a logistic function of $X$ and the outcome model is fitted as a linear function of $X$. To implement matching strategies and non-parametric local constant regression, we adopt $M = 5$, \bcol{the $L_2$ norm (or the euclidean distance) as the default distance metric},
and the bandwidth $h$ is chosen by using the  \texttt{dpill()} function in the \textsf{R} package \texttt{KernSmooth}.

\subsection{Simulation results}

We select several evaluation quantities as criteria and use them to assess our proposed methods, comparing them with some competitive methods.
All the simulation results are based on 1000 replications. Among the evaluation criteria, Bias, SD, and MSE are the Monte Carlo bias, standard deviation, and mean squared error over the 1000 simulations of the point estimates, respectively. \bcol{These comparisons of Bias, SD, and MSE across different methods are conducted with a sample size of $n = 2000$.}

\begin{table}[htbp]
  \centering
  \setlength{\tabcolsep}{1.7mm}
  \renewcommand\arraystretch{1.2}
  \caption{Comparison of Bias and SD of various methods for cases (C1)-(C3).}
    \begin{tabular}{cccccccccccccc}
    \toprule
    & & \multicolumn{2}{c}{MATCH} & \multicolumn{2}{c}{MATCH.bc} & \multicolumn{2}{c}{IPW} & \multicolumn{2}{c}{OR} & \multicolumn{2}{c}{AIPW} & \multicolumn{2}{c}{PSR} \\
Case&       z& Bias  & SD    & Bias  & SD    & Bias  & SD    & Bias  & SD    & Bias  & SD    & Bias  & SD \\
\hline
    & -0.4  & -0.009  & 0.119  & -0.008  & 0.113  & 0.021  & 0.276  & -0.272  & 0.124  & 0.002  & 0.175  & -0.098& 0.193
\\
& -0.2  & 0.007  & 0.114  & 0.007  & 0.108  & -0.003  & 0.272  & 0.029  & 0.092  & 0.001  & 0.165  & 0.238 & 0.159
\\
C1& 0     & 0.017  & 0.111  & 0.015  & 0.106  & -0.013  & 0.274  & 0.172  & 0.079  & 0.016  & 0.166  & 0.309 & 0.148
\\
& 0.2   & 0.020  & 0.112  & 0.019  & 0.108  & 0.008  & 0.282  & 0.154  & 0.094  & 0.022  & 0.171  & 0.344 & 0.153
\\
& 0.4   & -0.005  & 0.123  & -0.005  & 0.117  & 0.047  & 0.304  & -0.026  & 0.127  & -0.003  & 0.175  & 0.262 & 0.194
\\
    \midrule
    & -0.4  & -0.013  & 0.122  & -0.014  & 0.115  & 0.015  & 0.264  & -0.242  & 0.125  & -0.006  & 0.169  & 0.138 & 0.191
\\
& -0.2  & 0.007  & 0.110  & 0.007  & 0.105  & 0.000  & 0.270  & 0.021  & 0.090  & -0.002  & 0.164  & 0.246 & 0.150
\\
C2& 0     & 0.008  & 0.109  & 0.007  & 0.104  & -0.025  & 0.281  & 0.115  & 0.074  & 0.008  & 0.158  & 0.292 & 0.144
\\
& 0.2   & -0.003  & 0.108  & -0.003  & 0.103  & -0.024  & 0.285  & 0.062  & 0.091  & 0.001  & 0.162  & 0.288 & 0.157
\\
& 0.4   & -0.011  & 0.118  & -0.010  & 0.113  & 0.060  & 0.294  & 0.038  & 0.127  & -0.006  & 0.170  & 0.263 & 0.195
\\
    \midrule
    & -0.4  & 0.014  & 0.121  & 0.014  & 0.116  & 0.039  & 0.269  & 0.048  & 0.126  & 0.011  & 0.171  & 0.250 & 0.188
\\
& -0.2  & 0.004  & 0.115  & 0.003  & 0.109  & -0.014  & 0.279  & -0.076  & 0.093  & 0.003  & 0.168  & 0.204 & 0.152
\\
C3& 0     & -0.010  & 0.115  & -0.010  & 0.109  & -0.056  & 0.291  & -0.209  & 0.079  & -0.016  & 0.168  & 0.192 & 0.150
\\
& 0.2   & -0.023  & 0.115  & -0.023  & 0.109  & -0.032  & 0.302  & -0.147  & 0.096  & -0.019  & 0.171  & 0.203 & 0.162
\\
& 0.4   & 0.012  & 0.122  & 0.012  & 0.117  & 0.083  & 0.311  & 0.332  & 0.130  & -0.001  & 0.172  & 0.421 & 0.196
\\
    \bottomrule
    \end{tabular}
  \label{tab-comparison1_3}
\end{table}

\begin{table}[htbp]
  \centering
  \setlength{\tabcolsep}{1.7mm}
  \renewcommand\arraystretch{1.2}
  \caption{Comparison of Bias and SD of various methods for cases (C4)-(C6).}
    \begin{tabular}{cccccccccccccc}
    \toprule
    & & \multicolumn{2}{c}{MATCH} & \multicolumn{2}{c}{MATCH.bc} & \multicolumn{2}{c}{IPW} & \multicolumn{2}{c}{OR} & \multicolumn{2}{c}{AIPW} & \multicolumn{2}{c}{PSR} \\
Case&       z& Bias  & SD    & Bias  & SD    & Bias  & SD    & Bias  & SD    & Bias  & SD    & Bias  & SD \\
\hline
    & -0.4  & -0.044  & 0.152  & -0.029  & 0.147  & 0.294  & 0.329  & -0.290  & 0.163  & 0.026  & 0.222  & 0.058 & 0.280
\\
& -0.2  & 0.010  & 0.136  & 0.017  & 0.128  & -0.129  & 0.304  & 0.032  & 0.119  & 0.014  & 0.186  & 0.162 & 0.176
\\
C4& 0     & 0.014  & 0.137  & 0.014  & 0.129  & -0.413  & 0.333  & 0.190  & 0.101  & 0.038  & 0.189  & 0.210 & 0.169
\\
& 0.2   & 0.028  & 0.134  & 0.019  & 0.127  & -0.202  & 0.340  & 0.197  & 0.117  & 0.028  & 0.180  & 0.267 & 0.182
\\
& 0.4   & 0.033  & 0.152  & 0.014  & 0.146  & 0.644  & 0.352  & 0.039  & 0.154  & 0.035  & 0.207  & 0.229 & 0.286
\\
    \midrule
    & -0.4  & -0.039  & 0.149  & -0.024  & 0.145  & 0.205  & 0.300  & -0.257  & 0.154  & 0.011  & 0.199  & 0.060 & 0.264
\\
& -0.2  & 0.003  & 0.134  & 0.013  & 0.126  & -0.143  & 0.321  & 0.021  & 0.115  & 0.012  & 0.183  & 0.145 & 0.179
\\
C5& 0     & 0.012  & 0.132  & 0.010  & 0.123  & -0.399  & 0.318  & 0.133  & 0.100  & 0.027  & 0.174  & 0.215 & 0.160
\\
& 0.2   & 0.021  & 0.134  & 0.008  & 0.126  & -0.194  & 0.341  & 0.095  & 0.112  & 0.015  & 0.188  & 0.251 & 0.180
\\
& 0.4   & 0.024  & 0.152  & 0.008  & 0.148  & 0.631  & 0.360  & 0.083  & 0.154  & 0.000  & 0.210  & 0.217 & 0.290
\\
    \midrule
    & -0.4  & -0.020  & 0.153  & -0.004  & 0.148  & 0.268  & 0.313  & -0.006  & 0.161  & -0.013  & 0.209  & 0.132 & 0.285
\\
& -0.2  & -0.022  & 0.143  & -0.012  & 0.134  & -0.190  & 0.320  & -0.122  & 0.121  & -0.020  & 0.187  & 0.106 & 0.187
\\
C6& 0     & -0.012  & 0.135  & -0.014  & 0.126  & -0.553  & 0.336  & -0.242  & 0.100  & -0.050  & 0.179  & 0.151 & 0.158
\\
& 0.2   & -0.024  & 0.134  & -0.034  & 0.126  & -0.278  & 0.329  & -0.169  & 0.116  & -0.035  & 0.179  & 0.158 & 0.175
\\
& 0.4   & 0.033  & 0.151  & 0.015  & 0.146  & 0.626  & 0.345  & 0.322  & 0.161  & -0.059  & 0.215  & 0.269 & 0.281
\\
    \bottomrule
    \end{tabular}
  \label{tab-comparison4_6}
\end{table}

\begin{table}[htbp]
  \centering
  \setlength{\tabcolsep}{1.7mm}
  \renewcommand\arraystretch{1.2}
  \caption{Comparison of Bias and SD of various methods for cases (C7)-(C9).}
    \begin{tabular}{cccccccccccccc}
    \toprule
    & & \multicolumn{2}{c}{MATCH} & \multicolumn{2}{c}{MATCH.bc} & \multicolumn{2}{c}{IPW} & \multicolumn{2}{c}{OR} & \multicolumn{2}{c}{AIPW} & \multicolumn{2}{c}{PSR} \\
Case&       z& Bias  & SD    & Bias  & SD    & Bias  & SD    & Bias  & SD    & Bias  & SD    & Bias  & SD \\
\hline
    & -0.4  & -0.063  & 0.145  & -0.086  & 0.141  & -0.003  & 0.392  & -0.388  & 0.142  & -0.007  & 0.250  & -0.144 & 0.240
\\
& -0.2  & -0.025  & 0.118  & -0.019  & 0.113  & 0.025  & 0.274  & -0.055  & 0.103  & 0.009  & 0.174  & 0.009 & 0.157
\\
C7& 0     & -0.020  & 0.107  & -0.011  & 0.103  & 0.003  & 0.255  & 0.123  & 0.091  & 0.006  & 0.157  & 0.067 & 0.150
\\
& 0.2   & -0.013  & 0.123  & -0.005  & 0.118  & 0.008  & 0.306  & 0.140  & 0.116  & 0.015  & 0.189  & 0.096 & 0.186
\\
& 0.4   & 0.015  & 0.160  & 0.004  & 0.156  & -0.007  & 0.539  & -0.005  & 0.160  & 0.006  & 0.306  & -0.032 & 0.286
\\
    \midrule
    & -0.4  & -0.033  & 0.143  & -0.074  & 0.138  & 0.009  & 0.375  & -0.265  & 0.135  & 0.002  & 0.233  & -0.099 & 0.244
\\
& -0.2  & -0.011  & 0.108  & -0.010  & 0.104  & 0.016  & 0.259  & 0.001  & 0.099  & 0.017  & 0.164  & 0.032 & 0.157
\\
C8& 0     & -0.020  & 0.108  & -0.012  & 0.103  & -0.002  & 0.246  & 0.097  & 0.088  & 0.006  & 0.151  & 0.078 & 0.146
\\
& 0.2   & -0.025  & 0.124  & -0.018  & 0.119  & 0.007  & 0.324  & 0.047  & 0.113  & -0.004  & 0.190  & 0.067 & 0.194
\\
& 0.4   & 0.003  & 0.158  & -0.009  & 0.156  & -0.005  & 0.513  & 0.025  & 0.156  & -0.005  & 0.294  & -0.005 & 0.291
\\
    \midrule
    & -0.4  & 0.047  & 0.146  & 0.032  & 0.141  & 0.019  & 0.377  & 0.334  & 0.143  & 0.010  & 0.247  & 0.067 & 0.239
\\
& -0.2  & -0.016  & 0.114  & -0.009  & 0.110  & 0.010  & 0.271  & 0.105  & 0.101  & 0.004  & 0.172  & 0.041 & 0.162
\\
C9& 0     & -0.035  & 0.109  & -0.025  & 0.105  & -0.016  & 0.260  & -0.134  & 0.087  & -0.008  & 0.152  & 0.008 & 0.153
\\
& 0.2   & -0.043  & 0.121  & -0.035  & 0.116  & -0.025  & 0.322  & -0.177  & 0.109  & -0.022  & 0.186  & -0.062 & 0.196
\\
& 0.4   & 0.017  & 0.163  & 0.003  & 0.160  & 0.020  & 0.535  & 0.199  & 0.154  & 0.008  & 0.296  & 0.091 & 0.284
\\
    \bottomrule
    \end{tabular}
  \label{tab-comparison7_9}
\end{table}

Table \ref{tab-comparison1_3} summarizes the results for all six methods being compared in terms of Bias and SD for cases (C1)-(C3), where the propensity scores are not close to $0$ and $1$. \bcol{The Bias and SD of the GATE estimates are reported for five representative points $z = -0.4, -0.2, 0.0, 0.2,$ and $0.4$. Both Bias and SD are calculated as the Monte Carlo bias and standard deviation based on $1,000$ simulation replications of the point estimates.} For all three cases, the Bias of the MATCH, MATCH.bs, and weighting-based methods are small, and the SD of our proposed methods are smaller in comparison to the weighting-based methods. The OR and PSR methods are biased and not robust to model misspecification. Results for cases (C4)-(C6) are summarized in Table \ref{tab-comparison4_6}, where some propensity scores are close to $0$ and $1$. The MATCH and MATCH.bc methods demonstrate the robustness of model misspecification and extreme values in the propensity score, exhibiting small bias and SD. The IPW method shows poor performance in the presence of extreme propensity score values. Table \ref{tab-comparison7_9} shows the results for cases (C7)-(C9), which use different true propensity score models from cases 4-6 but still have some extreme propensity score values. As expected, MATCH and MATCH.bc still perform well. For all the cases, the proposed methods appear to be more robust in terms of BIAS and SD.

\begin{figure}
    \centering
    \includegraphics[width=0.7\linewidth]{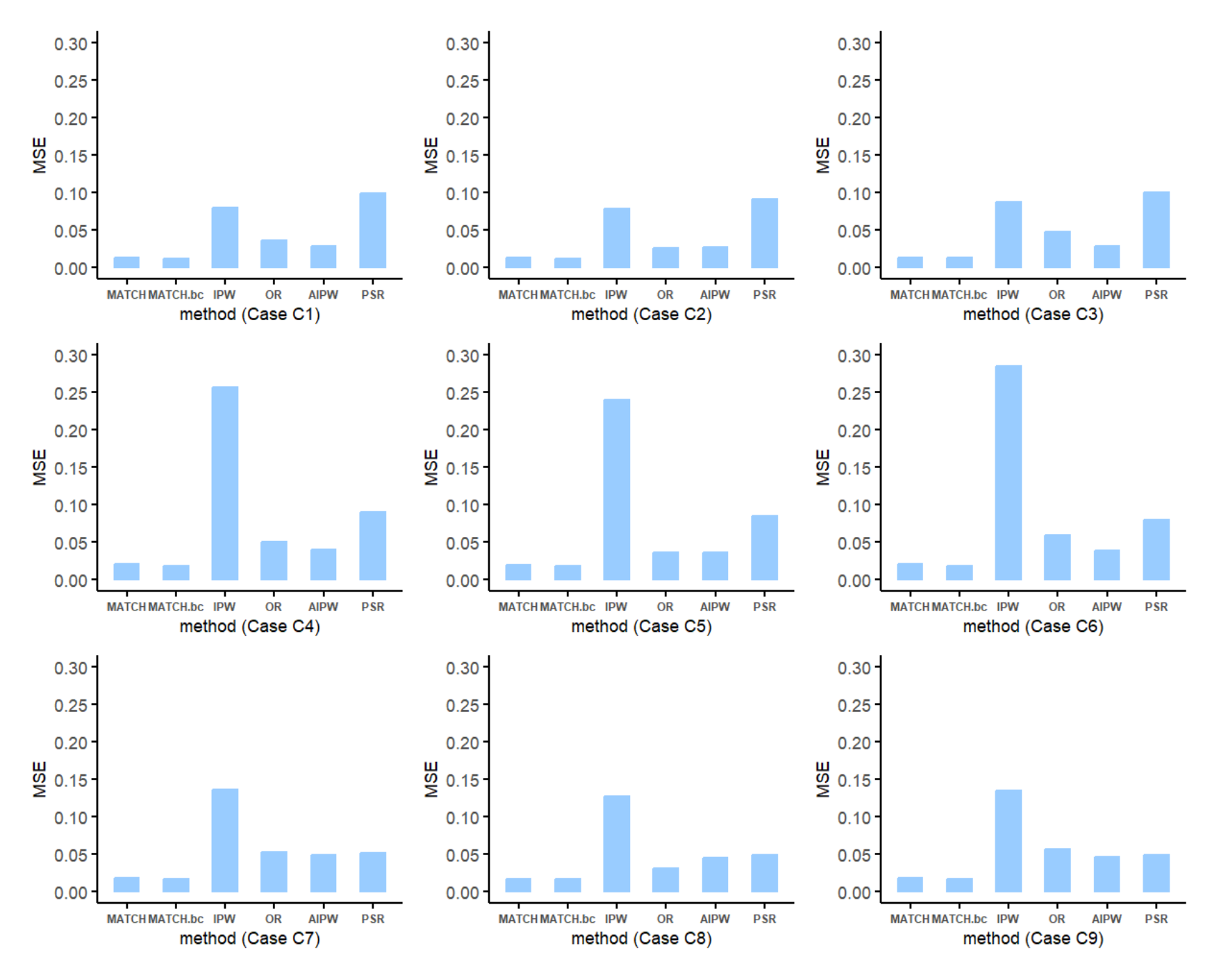}
    \caption{Comparison of MSE of various methods for cases (C1)-(C9). The competing methods include inverse probability weighted estimator (IPW); outcome regression (OR); augmented inverse probability weighting estimator (AIPW); propensity score regression estimator (PSR).}
    \label{fig-MSE}
\end{figure}

\bcol{In Figure \ref{fig-MSE}, for ease of presentation and comparison, we also calculate the MSE of the six competing methods. The MSE shown is the average of $\text{MSE}(z)$ for $z = -0.4$, $-0.2$, $0.0$, $0.2$, and $0.4$. Specifically,
The MSE shown in Figure 2 is
  \[  \text{MSE} =  \frac{  \text{MSE}(-0.4) + \text{MSE}(-0.2) + \text{MSE}(0.0) + \text{MSE}(0.2) + \text{MSE}(0.4) }{5}.      \] The specific MSE values for the six methods are provided in Table \ref{tab-MSE} in Appendix.} Compared to the other four methods, both proposed methods have the MSE values closer to $0$. This phenomenon is observed across all nine cases, indicating that MATCH and MATCH.bc exhibit favorable estimation performance. And as expected, MATCH.bc has a better performance than MATCH in most cases.

\begin{figure}
    \centering
    \includegraphics[width=0.7 \linewidth]{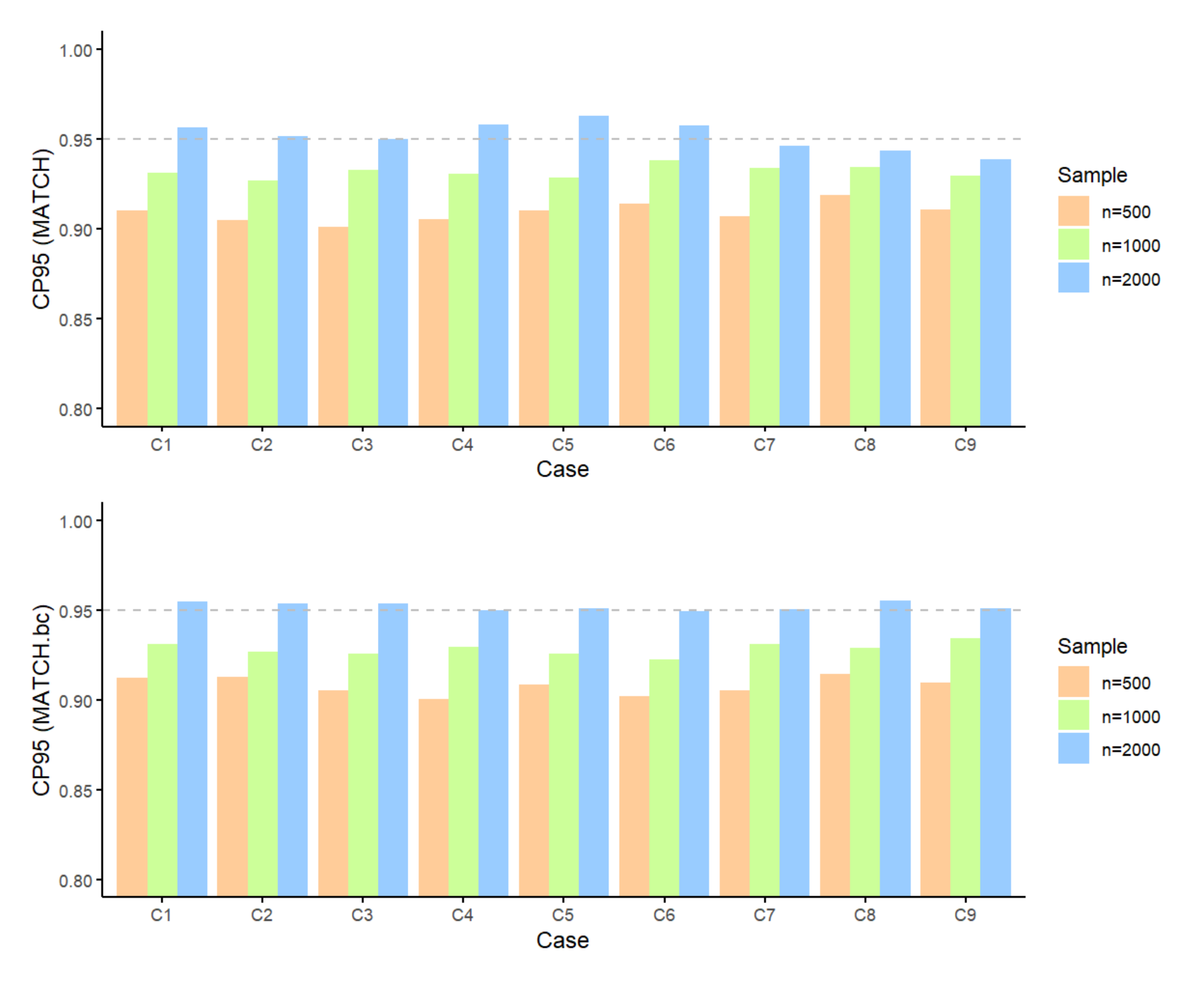}
    \caption{Average values of CP95 at five representative points based on 1000 simulations with sample size 500, 1000, 2000, where CP95 is the coverage proportions of the 95\% confidence intervals.}
    \label{fig-CP95}
\end{figure}

\bcol{Furthermore, we also evaluate the $95\%$ confidence interval coverage rates (CP95) for the proposed methods by employing the subsampling strategy.}
Specifically, for the nine cases, we compute the confidence interval coverage rates for GATE at five representative points of $Z= -0.4, -0.2, 0, 0.2, 0.4.$ We then take the average of these rates as the estimation of the $95\%$ confidence interval coverage rates. The sample sizes set in this simulation are $500$, $1000$, and $2000$. Figure \ref{fig-CP95} illustrates the results of the MATCH and MATCH.bc methods for cases (C1)-(C9), indicating that the CP95 values are approximately at the nominal level of $0.95$. Furthermore, the CP95 values gradually increase and converge towards the nominal level of $0.95$ with the increase in sample size.

\begin{figure}
    \centering
    \includegraphics[width=0.7\linewidth]{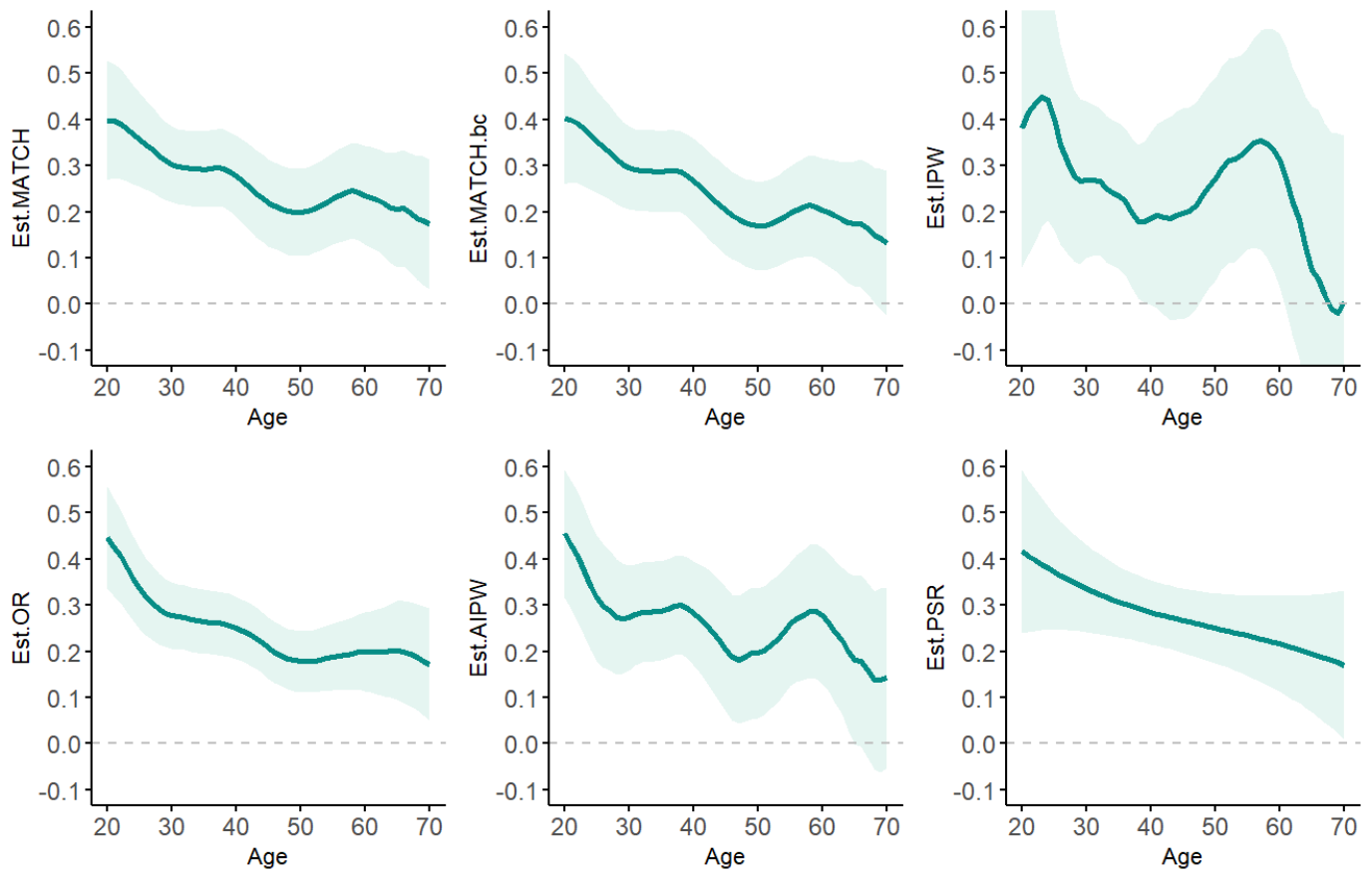}
    \caption{Estimated GATE and the associated 95\% CI for PASI 75 with \textbf{Age} based on six methods.}
    \label{fig-Application}
\end{figure}

\bcol{In the above simulation, we adopt the $L_2$-norm as the distance metric. To explore the sensitivity of our proposed  methods to the choice of distance metric, we conduct additional experiments by using three representative distance metrics. The associated numerical results are provided in Appendix \ref{App-C2}, which indicates that the proposed estimators perform robustly across various distance metrics.
}
\bcol{Moreover, the proposed methods rely on the strongly ignorability assumption. We further investigate the sensitivity of the proposed methods in the presence of missing confounders, and the corresponding numerical results are presented in Appendix~\ref{App-C3}, from which we find that the proposed methods still yield lower MSE than the competing approaches.}
\bcol{We also explore the impact of number of matches ($M$) for each unit. The associated numerical results are presented in Appendix~\ref{App-C4}, which indicates that the proposed estimators perform robustly for different numbers of matches.
}

\section{Application} \label{sec5}

Psoriasis, a chronic immune-mediated inflammatory disease \bcol{occurring worldwide~\cite{griffiths2007pathogenesis, griffiths2021psoriasis}.
This disease is marked by skin plaques with irregular borders and silvery scales, which can significantly diminish patients' quality of life~\cite{Ryan-etal2015} and are linked to an elevated risk of depression, anxiety, and even suicide~\cite{Kurd-etal2010}, imposing a substantial burden on society. Patients often prioritize rapid skin recovery and visible improvement~\cite{Blome-etal2016}. However, the chronic and relapsing nature of psoriasis, along with the need for sustained treatment, makes effective patient communication difficult. Globally, issues such as lack of treatment, inadequate therapy, and poor adherence continue to pose major challenges~\cite{Zhang2020}.}
Extensive research indicates that the primary drivers of psoriasis pathogenesis are proinflammatory cytokines such as tumor necrosis factor-alpha (TNF$\alpha$), interleukin-17 (IL-17), and interleukin-23 (IL-23)~\cite{park2005distinct}. The development of novel biologic therapies, including IL-17 inhibitors (secukinumab, ixekizumab, and brodalumab) and IL-23 inhibitors (guselkumab, risankizumab, and tildrakizumab), offers promising treatment options. A significant development occurred with the inclusion of secukinumab in China's national drug reimbursement list (NDRL) in 2020, which resulted in a substantial increase in its clinical use across the country.
Nevertheless, despite the escalating adoption of biologics therapies, evidence supporting their efficacy, particularly in cases of mild-to-moderate psoriasis, remains limited. Additionally, as the use of biologics therapies increases, concerns about their effective and appropriate use are also growing~\cite{chinese2020chinese}. \bcol{In the context of psoriasis treatment, substantial heterogeneity in treatment response to biologic therapies has been widely reported in the literature. Patient characteristics such as age, sex, disease duration, body mass index (BMI), comorbid conditions (e.g., psoriatic arthritis, metabolic syndrome), and prior treatment history can influence both the effectiveness and safety profiles of biologics ~\citep{Mahil2016_psoriasis, Pan2024_psoriasis}.} This study aims to address these issues by examining the heterogeneous treatment effects of biologics therapy compared to conventional therapy across various patient subpopulations. In doing so, it seeks to inform personalized treatment strategies for psoriasis and optimize patient care, \bcol{improving communications between patients and clinicians.}

\subsection{Data description}

The data for this study are obtained from the Psoriasis Center Data Platform (PCDP)\cite{yang-etal2023},  
administered by the National Clinical Research Center for Skin and Immune Disease. This platform provides data from 237 tertiary hospitals across approximately 100 cities in mainland China, encompassing a diverse patient cohort. The dataset analyzed in this section is restricted to individuals enrolled between September 2020 and September 2021, diagnosed with plaque psoriasis, and having completed at least one follow-up visit. Our analysis specifically focuses on patients treated with the IL-inhibitor biologics ($A = 1$), or with the conventional therapy ($A = 0$), such as topical drugs, systemic medications, or phototherapy. Patients who received other treatments are excluded from the analysis. The administration of biologics is categorized into an induction phase and a maintenance phase. We assess treatment effect heterogeneity during the maintenance phase, excluding patients with a follow-up period of less than four weeks. The final dataset for analysis comprises $2356$ samples, with 708 (30.05\%) receiving the biologics and 1648 (69.95\%) undergoing alternative therapies.

The psoriasis area and severity index (PASI) is the primary measure for assessing the clinical benefit of the IL-inhibitor biologics therapy.
When there is a $75\%$ or greater improvement in PASI, the outcome variable $Y$ (denoted as PASI 75) is coded as $1$; otherwise, it is coded as $0$.
\bcol{The covariates $X$ include patients' demographics, clinical characteristics, and their interactions. A detailed description is provided in Table \ref{tab-description} of the Appendix. Note that the proposed matching methods rely on the  unconfoundedness assumption (Assumption \ref{assump1}), and
there is always a risk of omitting important covariates~\cite{Imbens-Rubin-2015, Hernan-Robins-2020}.
To mitigate this, we incorporate as many observed variables as possible. Specifically, the 13 baseline covariates are selected based on prior clinical and empirical studies~\cite{yang-etal2023}. Building on this foundation, we expand the covariate set by explicitly including interaction terms, as shown in the last five rows of Table \ref{tab-description}.}
The subpopulation of interest is defined by age. Due to the scarcity of samples at extremely high age values, we cap age above $70$ at $70.$
We analyze the data using the six different methods, with implementation details consistent with those used in the simulation. In addition, we employ the subsampling method to estimate the confidence intervals.

\subsection{Results}

Figure \ref{fig-Application} displays the estimated GATE curves and corresponding 95\% CIs across the six different methods. The results of the proposed methods suggest a favorable effect of the biologics therapy compared to the conventional therapy, with observed treatment heterogeneity persisting across various subpopulations. Specifically, as age increases from 20 to 30, the relative advantage of biological therapy over conventional therapy slightly decreases. The effect stabilizes between ages 30 and 40. Between ages 40 and 60, the GATE follows a V-shaped curve, with a trough near age 50. Beyond age 60, the GATE tends to decrease, suggesting that advanced age is associated with diminished benefits from biologics therapy. The relative advantage of our proposed methods over age shows a trend similar to that of the OR method and the AIPW method but with smaller estimated standard errors. The PSR method shows a strictly monotonically decreasing trend in relative advantage with age, while the IPW method exhibits an increasing trend in older age groups (accompanied by large standard errors). Thus, the results of the IPW and PSR methods do not appear reasonable. The application results suggests the superiority of our proposed methods in the analysis of real-world data.


\section{Conclusion}\label{sec6}

\bcol{The group average treatment effects capture the heterogeneity of the average treatment effects within subgroups defined by covariates of interest.
While previous studies have primarily relied on weighting-based or regression-based estimation strategies for GATEs, these approaches face well-known limitations: weighting methods are often sensitive to extreme propensity scores—since they use the inverse of the propensity score as weights—leading to unstable estimates, whereas regression-based methods can suffer from bias due to model extrapolation, especially when there is limited overlap between the treated and control groups.
In this work, we propose two novel estimators: a matching-based estimator and a bias-corrected matching estimator. These methods are designed to address the limitations of existing approaches by avoiding reliance on inverse propensity score weights and improving covariate balance between treatment groups. As a result, they offer a more robust and interpretable framework for estimating GATEs. In addition, we rigorously analyze the large-sample properties of the proposed estimators, providing theoretical support for their validity and practical relevance.}

Despite possessing several desirable strengths, the proposed methods require the unconfoundedness assumption to hold. This necessitates that practitioners ensure the collection of all common confounders of the treatment and outcome variables. In future research, one possible strategy to relax this assumption is to introduce instrumental variables to address unobserved confounding. Additionally, we will consider extending our methods to estimate optimal individualized treatment rules, providing tailored treatment strategies based on specific variables of interest, such as age, gender, and other characteristics.

\bcol{It is noteworthy that while our proposed methods exhibit more stable estimation performance in cases where the true  propensity scores are close to zero or one, we still adhere to the standard positivity (overlap) assumption to ensure the identifiability of GATEs and do not consider scenarios in which this fundamental assumption is violated. Specifically, we assume $0< e(x) < 1$,  although some estimated propensity scores may be close to, or even exactly, $0$ or $1$ due to sampling error. In such cases, our estimators remain robust, whereas propensity score weighting methods either fail or suffer from high variance. We demonstrate this empirically in our simulations (see Table \ref{tab-comparison1.1} in Appendix \ref{App-C5}).}


\bibliographystyle{plainnat}
\bibliography{reference}



%

\newpage

  \begin{center}
\bf \Large
Supplementary Material
\end{center}

\setcounter{equation}{0}
\setcounter{section}{0}
\setcounter{figure}{0}
\setcounter{example}{0}
\setcounter{proposition}{0}
\setcounter{corollary}{0}
\setcounter{theorem}{0}
\setcounter{table}{0}
\setcounter{condition}{0}
\setcounter{lemma}{0}
\setcounter{remark}{0}

\renewcommand {\theproposition} {S\arabic{proposition}}
\renewcommand {\theexample} {S\arabic{example}}
\renewcommand {\thefigure} {S\arabic{figure}}
\renewcommand {\thetable} {S\arabic{table}}
\renewcommand {\theequation} {S\arabic{equation}}
\renewcommand {\thelemma} {S\arabic{lemma}}
\renewcommand {\thesection} {S\arabic{section}}
\renewcommand {\thetheorem} {S\arabic{theorem}}
\renewcommand {\thecorollary} {S\arabic{corollary}}
\renewcommand {\thecondition} {S\arabic{condition}}
\renewcommand {\thepage} {S\arabic{page}}
\renewcommand {\theremark} {S\arabic{remark}}

\setcounter{page}{1}

  \setcounter{equation}{0}
\renewcommand {\theequation} {S\arabic{equation}}
  \setcounter{lemma}{0}
\renewcommand {\thelemma} {S\arabic{lemma}}
   \setcounter{definition}{0}
\renewcommand {\thedefinition} {S\arabic{definition}}
   \setcounter{example}{0}
\renewcommand {\theexample} {S\arabic{example}}
   \setcounter{proposition}{0}
\renewcommand {\theproposition} {S\arabic{proposition}}
   \setcounter{corollary}{0}
\renewcommand {\thecorollary} {S\arabic{corollary}}

\section{Proofs in Section 3.1  \label{app-a}}

\subsection{Proof of Theorem 1}

\begin{proof}[Proof of Theorem 1]

Let $K_h(Z_i - z) =  h^{-1} K( \frac{Z_{i} - z}{h})$,  $ \bX$ and $\bA$ be the matrix and vector with $i$-th row equal to $ X_i^\top$ and $A_i$, respectively.
We shall decompose the difference between $\hat \tau(z)^{match}$ and $\tau(z)$ as follows,
\begin{align*}
		     \hat \tau(z)^{match} - \tau(z)
		     	={}& \frac{  N^{-1} \sum_{i=1}^{N} K_h(Z_{i} - z) \{ \hat Y_{i}^1- \hat Y_{i}^0  \} }   {  N^{-1} \sum_{i=1}^{N} K_h(Z_{i} - z)  } - \tau(z), \notag  \\
		     	={}&  \frac{ U_{1N}(z) + U_{2N}(z) + U_{3 N}(z) +  U_{4N}(z) }{  \hat f_N(z)  },  \label{eq-s1}
	\end{align*}
where
    \begin{align*}
                 U_{1N}(z) ={}&  N^{-1} \sum_{i=1}^{N} K_h(Z_{i} - z) \cdot (\tau( X_i) -  \tau(Z_i) ), \\
                     U_{2N}(z) ={}&   N^{-1} \sum_{i=1}^{N} K_h(Z_{i} - z)  \cdot (  \tau(Z_i) - \tau(z)),\\
                  U_{3N}(z)  ={}& N^{-1} \sum_{i=1}^{N} K_h(Z_{i} - z)  \Big [  \E( \hat Y_{i}^1 - \hat Y_{i}^0 \mid   \bX, \bA ) - \tau( X_i) \Big ], \\
	  U_{4N}(z)  ={}&  N^{-1} \sum_{i=1}^{N} K_h(Z_{i} - z)  \Big [ \hat Y_{i}^1- \hat Y_{i}^0  -  \E( \hat Y_{i}^1 - \hat Y_{i}^0 \mid  \bX, \bA )  \Big ],\\
             	  \hat f_N(z)  ={}&  N^{-1} \sum_{i=1}^{N} K_h(Z_{i} - z).
    \end{align*}		
The first term $U_{1N}(z)$ represents the deviation of
the true average conditional causal effects conditioning on random $ X_i$
from the true average conditional causal effects conditioning on the random $Z_i$.
The second term $U_{2N}(z)$
represents the deviation of the true average conditional causal effects conditioning on random $Z_i$
from the true average conditional causal effects conditioning on the fixed value of $z$.
The third term $U_{3N}(z)$ represents the deviation of the estimated average conditional causal effects conditioning on random $ X_i$ from the true average conditional causal effects conditioning on the random $ X_i$. The fourth term $U_{4N}(z)$ represents the remaining error terms that cannot be accounted for by non-parametric regression.

The denominator $\hat f_N(z)$ denotes the standard kernel density estimator, and under Assumptions \ref{assum_3}(i) and \ref{assum_3}(ii), we have (see Section 1.9 of Li and Racine \cite{Li-Racine-2007})
    \begin{equation}  \label{eq-s2}
        \hat f_N(z) - f(z)  = O_p(\frac{1}{\sqrt{Nh}} + h^2).
    \end{equation}

Below, we shall prove the four numerators.
We start with $U_{1N}(z)$. It is in fact a variance induced by kernel regression.  Since $\E[\tau(  X_i)| Z_i] = \tau(Z_i)$, the expectation of $U_{1N}(z)$ is zero. Thus, its variance is given by
    \begin{align*}
        \text{Var}(\sqrt{Nh} U_{1N}(z)) ={}& h \cdot \text{Var} \Big [   K_h(Z_{i} - z) \cdot ( \tau( X_i) -  \tau(Z_i)) \Big ] \\
        ={}& h \cdot \E \Big [ \frac{1}{h^2} K^2(\frac{Z_i - z}{h}) \cdot ( \tau( X_i) -  \tau(Z_i))^2       \Big ]  \\
        ={}&   \frac{1}{h} \cdot \E \Big [ K^2(\frac{Z_i - z}{h}) \cdot ( \tau( X_i) -  \tau(Z_i))^2  \Big ]   \\
        ={}&   \frac{1}{h} \cdot \E \Big [ K^2(\frac{Z_i - z}{h}) \cdot \E [ (\tau( X_i) -  \tau(Z_i))^2  | Z_i ]     \Big ] \\
        ={}&   \frac{1}{h} \cdot \int  K^2(\frac{u - z}{h}) \cdot \E [ ( \tau( X_i) -  \tau(Z_i))^2  | Z_i = u]  f(u) du \\
        ={}&   \int   K^2(t) \cdot \E \Big [ ( \tau( X) -  \tau(Z))^2  | Z = z + ht\Big ]  f(z + ht) dt \\
        ={}& \int K^2(t)dt \cdot \text{Var}(\tau( X) | Z =z) f(z + ht) dt,\\
        ={}& \int K^2(t)dt \cdot \text{Var}(\tau( X) | Z =z) \cdot f(z)  + O(h),
    \end{align*}
where the fourth equality follows from the law of iterated expectations, the sixth equality is due to the integral transformation, $u = z + ht$, the second last equality follows that $\E [ \{ \tau( X) -  \tau(Z)\}^2  | Z_i = z + ht] $ around $z$ under Assumptions \ref{assum_3}(i) and \ref{assum_3}(iii), and the last equality follows the Taylor expansion of $f(z + ht)$.
Therefore,
  \begin{equation}
         U_{1N} = O_p(1/\sqrt{Nh}).
  \end{equation}

We proceed with $U_{2N}(z)$. $U_{2N}(z)$ represents the bias induced by kernel regression.  The expectation of $U_{2N}(z)$ is
    \begin{align*}
        \E[U_{2N}(z)] ={}& \E \Big [ \frac{1}{h} \cdot K(\frac{Z_{i} - z}{h}) \cdot ( \tau(Z_i) - \tau(z)) ] \\
        ={}&  \int K(t) \cdot (\tau(z + ht) - \tau(z) ) f(z + ht) dt \\
        ={}&  \int K(t) \cdot \Big (\tau'(z) ht +  \frac{1}{2} \tau''(t) h^2 t^2 + o(h^2)\Big ) \cdot \Big ( f(z) + f'(z) h t + \frac{1}{2} f''(z) h^2 t ^2 + o(h^2) \Big ) dt \\
        ={}& \int K(t) t^2 dt \cdot \Big (\tau'(z) f'(z) h^2 + \frac{1}{2} \tau''(z) f(z) h^2 \Big ) + O(h^3).
    \end{align*}

Identical to the analysis of $U_{1n}(z)$, the variance of $U_{2N}(z)$ is $O_p(h^2/(Nh))$. Thus,
    \begin{equation}
        U_{2N}(z) = \int K(t) t^2 dt \cdot \Big ( \tau'(z) f'(z)  + \frac{1}{2} \tau''(z) f(z) \Big )\cdot h^2 + O_p(h/\sqrt{nh} + h^3 ).
    \end{equation}

We then proceed with $U_{3N}(z)$. $U_{3N}(z)$ represents the bias induced by matching estimator. Note that for $A_i = 1$,
    \begin{align*}
     \E( \hat Y_{i}^1 - \hat Y_{i}^0 \mid  \bX, \bA ) - \tau( X_i)
 ={}& \E[ Y_{i}^1 -  \frac 1 M \sum_{j \in \cJ_{M}(i)} Y_{j} \mid  \bX, \bA  ]- \tau( X_i)   \\
 ={}& \mu_1( X_i) -  \frac 1 M \sum_{j \in \cJ_{M}(i)} \E[ Y_{j} \mid  \bX, \bA  ]      -  ( \mu_1( X_i) -   \mu_0( X_i) ) \\
 ={}&  \mu_0( X_i) - \frac 1 M \sum_{m=1}^M \E[ Y_{l_m(i)} \mid  \bX, \bA  ] \\
 ={}&  \mu_0( X_i) - \frac 1 M \sum_{m=1}^M \E[ Y_{l_m(i)}^0  \mid  \bX, \bA  ]  \\
 ={}&  \mu_0( X_i) - \frac 1 M \sum_{m=1}^M \mu_{0}( X_{l_m(i)}).
    \end{align*}
Similarly, for $A_i = 0$, we have
    $\E( \hat Y_{i}^1 - \hat Y_{i}^0 \mid  \bX, \bA ) - \tau( X_i)
 = \frac 1 M \sum_{m=1}^M \mu_{1}( X_{l_m(i)}) -   \mu_1( X_i).$

 Therefore, $U_{3N}(z)$ can be written as
    \begin{align*}
	  U_{3N}(z)  ={}& N^{-1} \sum_{i=1}^{N} K_h(Z_{i} - z) \cdot \Big [  \E( \hat Y_{i}^1 - \hat Y_{i}^0 \mid \bX, \bA ) - \tau(  X_i) \Big ] \\
	  		={}&   N^{-1} \sum_{i=1}^{N} K_h(Z_{i} - z) \cdot  (2 A_i - 1) \left [  \frac 1 M \sum_{m=1}^M \{ \mu_{1- A_i}( X_i) - \mu_{1- A_i}( X_{l_m(i)})    \}      \right ].
    \end{align*}	

According to the Lemma 2 in Abadie and Imbens \cite{Abadie-Imbens-2006}, under the unconfoundednes Assumption  and Assumption \ref{assum_2}, $\sup_{ X_i \in  \mathcal{X}} | \{ \mu_{1- A_i}( X_i) - \mu_{1- A_i}( X_{j_m(i)})  | = O_p(N^{-1/p})$. This implies that
\begin{equation}
U_{3N}(z)  = O_p(N^{-1/p}).
\end{equation}

Finally, $U_{4N}(z)$ represents the variance induced by the remaining errors that cannot be accounted for by non-parametric regression.
Again, note that for $A_i = 1$,
\begin{align*}
   & \hat Y_{i}^1- \hat Y_{i}^0  -  \E[ \hat Y_{i}^1 - \hat Y_{i}^0 \mid   \bX, \bA ]   \\
  ={}&  Y_{i} -  \frac 1 M \sum_{j \in \cJ_{M}(i)} Y_{j} -  \E \Big [ Y_{i} -  \frac 1 M \sum_{j \in \cJ_{M}(i)} Y_{j}  \mid   \bX, \bA \Big ]  \\
  ={}&  Y_{i} - \E( Y_{i} \mid   \bX, \bA )  - \Big [  \frac 1 M \sum_{j \in \cJ_{M}(i)} \Big ( Y_{j}  -  \E( Y_{j} \mid   \bX, \bA   )\Big ) \Big ] \\
  ={}&  \epsilon_{i} -   \frac 1 M \sum_{j \in \cJ_{M}(i)}  \epsilon_{j},
\end{align*}
and for $A_i = 0$,
\begin{align*}
    \hat Y_{i}^1- \hat Y_{i}^0  -  \E[ \hat Y_{i}^1 - \hat Y_{i}^0 \mid   \bX, \bA ]   =  \frac 1 M \sum_{j \in \cJ_{M}(i)}  \epsilon_{j} - \epsilon_{i}.
  \end{align*}
This implies that
    \begin{align*}
        	U_{4N}(z)  ={}&  N^{-1} \sum_{i=1}^{N} K_h(Z_{i} - z) \cdot \Big [ \hat Y_{i}^1- \hat Y_{i}^0  -  \E\{ \hat Y_{i}^1 - \hat Y_{i}^0 \mid  \bX, \bA \}  \Big ] \\
		 ={}&  N^{-1} \sum_{i=1}^{N} K_h(Z_{i} - z) \cdot (2 A_i - 1) \cdot \Big ( \epsilon_{i} - \frac 1 M \sum_{j \in \cJ_{M}(i)}  \epsilon_{j} \Big ).
    \end{align*}

   Thus, $U_{4N}(z)$ has zero mean. To obtain the variance, we divide $U_{1N}(z)$ into two terms, i.e.,
     \begin{align*}
            U_{4N}(z) ={}& U_{4N1}(z) - U_{4N2}(z),
    \end{align*}
   where
  \begin{align*}
      U_{4N1}(z)  ={}&  N^{-1} \sum_{i=1}^{N} K_h(Z_{i} - z)  A_i \Big ( \epsilon_{i} - \frac 1 M \sum_{j \in \cJ_{M}(i)}  \epsilon_{j} \Big ) \\
       U_{4N2}(z)  ={}& N^{-1} \sum_{i=1}^{N} K_h(Z_{i} - z) (1-A_i) \Big ( \epsilon_{i} - \frac 1 M \sum_{j \in \cJ_{M}(i)}  \epsilon_{j} \Big).
  \end{align*}

Since $\text{Cov}( U_{4N1}(z),  U_{4N2}(z) ) = 0$, we can compute the variance of $\text{Var}(U_{4N1}(z))$ and $\text{Var}(U_{4N2}(z))$ separately. For
 $\text{Var}(U_{4N1}(z))$, observe that

 \begin{align*} \label{eeq:U41n}
U_{4N1}(z)  =  N^{-1} \sum_{i=1}^{N} K_h(Z_{i} - z) A_i  \epsilon_{i} -  N^{-1} \sum_{i=1}^{N} \Big [ K_h(Z_{i} - z) A_i  \cdot \frac 1 M \sum_{j \in \cJ_{M}(i)}  \epsilon_{j} \Big ]
 \end{align*}

 \begin{align*}
       & \text{Cov}\left( K_h(Z_{i} - z) A_i  \epsilon_{i}, K_h(Z_{i} - z) A_i  \cdot \frac 1 M \sum_{j \in \cJ_{M}(i)}  \epsilon_{j} \right )  \\
    ={}& \E \Big [  K_h^2(Z_{i} - z) A_i ^2 \epsilon_{i} \cdot \Big ( \frac 1 M \sum_{j \in \cJ_{M}(i)}  \epsilon_{j} \Big ) \Big ]    \\
    ={}& \E[  K_h^2(Z_{i} - z) A_i ^2 \epsilon_{i} ] \cdot \E[ \frac 1 M \sum_{j \in \cJ_{M}(i)}  \epsilon_{j} ] \\
    ={}& 0.
 \end{align*}
Thus, we can estimate the variance for the two terms in $U_{4N1}(z)$ separately. Following an identical analysis to $U_{1N}(z)$, we have the variance of the first term $\text{Var}\Big ( N^{-1} \sum_{i=1}^{N} K_h(Z_{i} - z) A_i  \epsilon_{i} \Big ) = O_p(1/(Nh))$. The variance of the second term is given by
\begin{align*}
 &  \text{Var}\left [  N^{-1} \sum_{i=1}^{N} \left ( K_h(Z_{i} - z) A_i  \cdot \frac 1 M \sum_{j \in \cJ_{M}(i)} \epsilon_{j} \right ) \right ] \\
    ={}& \frac{1}{N^2} \Biggl [ \sum_{i=1}^N \text{Var}\Big (  K_h(Z_{i} - z) A_i  \cdot \frac 1 M \sum_{j \in \cJ_{M}(i)} \epsilon_{j}  \Big )  \\
   +  \sum_{1\leq i\leq N; 1\leq k\leq N; i \neq k} {}& \text{Cov}\Big ( K_h(Z_{i} - z) A_i  \cdot \frac 1 M \sum_{j \in \cJ_{M}(i)} \epsilon_{j},  K_h(Z_{k} - z) A_k  \cdot \frac 1 M \sum_{s \in \cJ_{M}(k)} \epsilon_{s} \Big ) \Biggr   ].
\end{align*}
 Let $K_M(i)$ denote the number of times unit $i$ is used as a match, i.e.,  $K_M(i) = \sum_{l=1}^N I\{i \in \cJ_M(l)\}.$ By the Lemma 3 of Abadie and Imbens\cite{Abadie-Imbens-2006},  $K_M(i) = O_p(1)$ and $\E[(KM(i))^q]$  is bounded uniformly for any constant $q > 0$.
 This means that there are only a finite number of terms in
 $ \sum_{1\leq i\leq N; 1\leq k\leq N; i \neq k} \text{Cov}\Big ( K_h(Z_{i} - z) A_i  \cdot \frac 1 M \sum_{j \in \cJ_{M}(i)} \epsilon_{j},  K_h(Z_{k} - z) A_k  \cdot \frac 1 M \sum_{s \in \cJ_{M}(k)} \epsilon_{s} \Big )$
 that are not equal to zero. Therefore, there exists a constant $C$ such that
 \begin{align*}
      & \text{Var}\left [  N^{-1} \sum_{i=1}^{N} \left ( K_h(Z_{i} - z) A_i  \cdot \frac 1 M \sum_{j \in \cJ_{M}(i)} \epsilon_{j} \right ) \right ]    \\
     \leq{}&   \frac{C}{N^2} \Biggl [ \sum_{i=1}^N \text{Var}\Big (  K_h(Z_{i} - z) A_i  \cdot \frac 1 M \sum_{j \in \cJ_{M}(i)} \epsilon_{j}  \Big ) \Biggr ] = O_p(\frac{1}{Nh}).
 \end{align*}
As such, $\text{Var}(U_{4N1}(z)) = O_p(1/(Nh))$. Similarly, $\text{Var}(U_{4N2}(z)) = O_p(1/(Nh))$. This leads to the conclusion that
\begin{equation}  \label{eq-s6}
    U_{4N}(z) = O_p(1/\sqrt{Nh}).
\end{equation}

Finally, the conclusion of Theorem 1 follows immediately from the equations (\ref{eq-s2})--(\ref{eq-s6}).  The rate of convergence $ O_p(N^{-1/p} + h^2 + 1/\sqrt{Nh} )$ includes two terms, where $O_p(N^{-1/p})$ comes from matching and $O_p(h^2 + 1/\sqrt{Nh})$ comes from nonparametric kernel regression.

\end{proof}

\subsection{Proof of Proposition 1}


\begin{proof}[Proof of Proposition 1] We first prove the Proposition \ref{prop1}(a). For simplification, let $W_i = K_h(Z_i - z) \Big /\sum_{j=1}^N K_h(Z_j - z)$ be the weight.  Then
 		   \[  \hat \tau(z)^{match}
		     	= \frac{ \sum_{i=1}^{N} K_h(Z_{i} - z) \{ \hat Y_{i}^1- \hat Y_{i}^0  \} }   {   \sum_{i=1}^{N} K_h(Z_{i} - z)  }  = \sum_{i=1}^N W_i  \{ \hat Y_{i}^1- \hat Y_{i}^0  \}.  \]
If $A_i =1$,  we have
     \begin{align*}
       W_i  \{ \hat Y_{i}^1- \hat Y_{i}^0  \}
        ={}&  W_i  (Y_{i} -  \frac 1 M \sum_{j \in \cJ_{M}(i)} Y_{j} ) \\
        ={}&
       W_i  Y_{i} -  \frac 1 M \sum_{j \in \cJ_{M}(i)} W_j Y_{j}   +  \frac 1 M \sum_{j \in \cJ_{M}(i)} (W_j - W_i) Y_{j}.
       \end{align*}
Since $\cJ_{M}(i)$ is the first $M$ matches for unit $i$, we have $\sup_i |W_j - W_i| = O_{\P}(N^{-1/p})$ by the Lemma 2 of Abadie and Imbens \cite{Abadie-Imbens-2006}. Likewise, for $A_i = 0$, we have
       \begin{align*}
       W_i  \{ \hat Y_{i}^1- \hat Y_{i}^0  \} =
          \frac 1 M \sum_{j \in \cJ_{M}(i)} W_j Y_{j} -  W_i  Y_{i}   +  O_{\P}(N^{-1/p}).
       \end{align*}
Thus,
    \[  \hat \tau(z)^{match}  =  \sum_{i=1}^N (2A_i - 1) \left ( W_i + \frac{ K_M(i) }{M}  W_i   \right )  Y_i +  O_{\P}(N^{-1/p}),   \]
 which implie the conclusion of  Proposition \ref{prop1}(a).

  Next, we show the Proposition \ref{prop1}(b). If $f(X=x \mid A=0)$ and $f(X=x \mid A=1)$ satisfies the Assumption 4.3 of Lin et al. \cite{Lin-etal2023} and $M = c N^{2/(2+p)}$ for some constant $c > 0$, then by a similar proof of Proposition \ref{prop1}(a),
we have
          \begin{equation} \label{eq-A7}  \hat \tau(z)^{match}  = \frac{ \sum_{i=1}^{N} K( \frac{Z_{i} - z}{h}) \left \{  A_i Y_i \cdot \left (1+ \dfrac{K_M(i)}{M} \right )  - (1-A_i)Y_i \cdot\left (1+ \dfrac{K_M(i)}{M} \right )    \right \} }  {  \sum_{i=1}^{N} K( \frac{Z_{i} - z}{h})} +  O_{\P}(N^{-1/(2+p)}),   \end{equation}
by using the Theorem 4.4 of Lin et al.\cite{Lin-etal2023}. In addition, according to the findings of Lin et al. \cite{Lin-etal2023} (Section 5),
   \begin{align}  \label{eq-A8}
  \begin{cases}
            1+ \frac{K_M(i)}{M}  ={}&  \frac{1}{ \pi(X_i)} +  O_{\P}(  N^{-1/(2+p)} ), \quad \text{if $A_i = 1$} \\
            1+ \frac{K_M(i)}{M}  ={}&  \frac{1}{1- \pi(X_i)} +  O_{\P}(  N^{-1/(2+p)} ), \quad \text{if $A_i = 0$}.
     \end{cases}
     \end{align}
   Then Proposition \ref{prop1}(b) follows immediately by combining (\ref{eq-A7}) and (\ref{eq-A8}).

\end{proof}

\section{Proofs in Section 3.2  \label{app-b}}


\subsection{Proofs of Propositions 2 and 3}
\begin{proof}[Proof of Proposition 2]  By the Theorem 5.1 of Lin et al. \cite{Lin-etal2023}, when  condition  (i) holds,  $\sup_{ X_i \in  \mathcal{X}} || X_i - X_{j_m(i)}  || = o_{\P}(1)$,
          \begin{equation*}  \hat \tau(z)^{match}  = \frac{ \sum_{i=1}^{N} K( \frac{Z_{i} - z}{h}) \left \{  A_i Y_i \cdot \left (1+ \dfrac{K_M(i)}{M} \right )  - (1-A_i)Y_i \cdot\left (1+ \dfrac{K_M(i)}{M} \right )    \right \} }  {  \sum_{i=1}^{N} K( \frac{Z_{i} - z}{h})} +  o_{\P}(1),   \end{equation*}
and
\[ \begin{cases}
            1+ K_M(i)/M  =  1/\pi(X_i) +  o_{\P}(1) \text{ if $A_i = 1$}, \\
            1+ K_M(i)/M =  1/(1- \pi(X_i)) +  o_{\P}(1) \text{ if $A_i = 0$}.
            \end{cases}
            \]
This implies the consistency of $\hat \tau(z)^{mathch}$ holds by the consistency of the IPW estimator.  In addition,
             the consistency of $\hat \tau^{bc}$ is derived by noting that $\hat \tau(z)^{bc} = \hat \tau(z)^{mathch} + o_{\P}(1)$.

 When condition (ii) holds, note that for $A_i = 0$,
			\begin{align} \label{eq-B9}
				\tilde Y_{i}^1 ={}&
				 \hat \mu_1(X_i)  +  \Big \{ \frac 1 M \sum_{j \in \cJ_{M}(i)} (Y_{j}^1 - \hat \mu_{1}(X_{j}) ) \Big \},
			\end{align}
Let  $ \bX$ be the matrix with $i$-th row equal to $ X_i^\top$. Since $\E[\hat Y_i^1 | \bX ] = \frac 1 M \sum_{j \in \cJ_{M}(i)} \mu_{1}(X_{j})$, if the estimated regression function is a uniformaly consistent estimator to the true regression function, then
       \[ \frac{ \sum_{i=1}^{N} K_h(Z_{i} - z) (1-A_i)  \Big \{ \frac 1 M \sum_{j \in \cJ_{M}(i)} (Y_{j}^1 - \hat \mu_{1}(X_{j}) )\Big \} }   {   \sum_{i=1}^{N} K_h(Z_{i} - z)  }  =  o_{\P}(1).   \]
   For $A_i = 1$, we have similar conclusion.  Thus, according to equation \eqref{eq-B9},
    the consistency of $\hat \tau^{bcn}$ is obtained by the consistency of $\hat \mu_a(X)$.

\end{proof}

\medskip
\begin{proof}[Proof of Proposition 3]
We first decompose $\hat \tau(z)^{bc} $ as follows,
	  \begin{align*}
	  &   \hat \tau(z)^{bc}  \\
	  ={}&   \frac{ \sum_{i=1}^{N}  K_h(Z_{i} - z)  \{ \tilde Y_{i}^1- \tilde Y_{i}^0  \} }   {  \sum_{i=1}^{N} K_h(Z_{i} - z)  }   \\
	    ={}&      \frac{ \sum_{i=1}^{N}  K_h(Z_{i} - z) A_i \{ Y_{i}^1-  \frac 1 M \sum_{j \in \cJ_{M}(i)} \{ Y_{j}^0  + \hat \mu_{0}(X_{i}) - \hat \mu_{0}(X_{j})  \} \} }   {  \sum_{i=1}^{N} K_h(Z_{i} - z)  } \\  &+    \frac{ \sum_{i=1}^{N}  K_h(Z_{i} - z) (1-A_i) \{ \frac 1 M \sum_{j \in \cJ_{M}(i)} \{ Y_{j}^1 + \hat \mu_{1}(X_{i}) - \hat \mu_{1}(X_{j})  \}   - Y_i^0 \}  }   {  \sum_{i=1}^{N} K_h(Z_{i} - z)  }     \\
	    ={}& \frac{ \sum_{i=1}^{N}  K_h(Z_{i} - z) A_i \{ (Y_{i}^1-  \hat \mu_1(X_i))  + (\hat \mu_1(X_i) - \hat \mu_0(X_i)) -   \frac 1 M \sum_{j \in \cJ_{M}(i)} \{ Y_{j}^0  - \hat \mu_{0}(X_{j})  \} \} }   {  \sum_{i=1}^{N} K_h(Z_{i} - z)  } \\
	    {}& +    \frac{ \sum_{i=1}^{N}  K_h(Z_{i} - z) (1-A_i) \{ \frac 1 M \sum_{j \in \cJ_{M}(i)} \{ Y_{j}^1  - \hat \mu_{1}(X_{j})  \}   + ( \hat \mu_{1}(X_{i}) -  \hat \mu_{0}(X_{i})) +  (\hat \mu_{0}(X_{i})  - Y_i^0) \}  }   {  \sum_{i=1}^{N} K_h(Z_{i} - z)  }  \\
	   ={}&   \frac{ \sum_{i=1}^{N} K_h(Z_{i} - z) \Big \{  A_i (Y_i - \hat \mu_1(X_i))  - (1-A_i)(Y_i - \hat \mu_0(X_i))   +  (\hat \mu_1(X_i) - \hat \mu_0(X_i) ) \Big \} }  {  \sum_{i=1}^{N} K_h(Z_{i} - z) }   \\
	   {}&  -  \frac{ \sum_{i=1}^{N}  K_h(Z_{i} - z) A_i \{ \frac 1 M \sum_{j \in \cJ_{M}(i)} \{ Y_{j}^0  - \hat \mu_{0}(X_{j})  \} \} }   {  \sum_{i=1}^{N} K_h(Z_{i} - z)  } \\ &+     \frac{ \sum_{i=1}^{N}  K_h(Z_{i} - z) (1-A_i) \{ \frac 1 M \sum_{j \in \cJ_{M}(i)} \{ Y_{j}^1  - \hat \mu_{1}(X_{j})  \}  }   {  \sum_{i=1}^{N} K_h(Z_{i} - z)  }  \\
	   ={}&   \frac{ \sum_{i=1}^{N} K( \frac{Z_{i} - z}{h}) \Big \{  A_i (Y_i - \hat \mu_1(X_i)) \cdot (1+ \dfrac{K_M(i)}{M})  - (1-A_i)(Y_i - \hat \mu_0(X_i)) \cdot (1+ \dfrac{K_M(i)}{M})    +  (\hat \mu_1(X_i) - \hat \mu_0(X_i) ) \Big \} }  {  \sum_{i=1}^{N} K( \frac{Z_{i} - z}{h})} \\
	   {}& +  O_{\P}\Big (  \sup_{j \in \cJ_M(i)} || K(\frac{Z_{i} - z}{h}) -  K(\frac{Z_{j} - z}{h})   ||  \cdot ( || \hat \mu_0 - \mu_0  ||_{\infty} +  || \hat \mu_1 - \mu_1  ||_{\infty} ) \Big ),
	   	    \end{align*}
where $j \in \cJ_M(i)$.    By the Theorem 4.4 of Lin et al.\cite{Lin-etal2023}, $\sup_{j \in \cJ_M(i)} || K(\frac{Z_{i} - z}{h}) -  K(\frac{Z_{j} - z}{h})   || = O_{\P}(N^{-1/(2+p)})$, combining it with equation \eqref{eq-A8} yields the conclusion.

	\end{proof}


\subsection{Proof of Theorem 2}

\begin{proof}[Proof of Theorem 2]  By the Proposition \ref{prop3}, 	
         \begin{align*}
         &    \hat \tau(z)^{bc} \\
          ={}& \sum_{i=1}^{N} K( \frac{Z_{i} - z}{h}) \Big \{  \dfrac{A_i (Y_i - \hat \mu_1(X_i))}{ \pi(X_i)} - \dfrac{(1-A_i)(Y_i - \hat \mu_0(X_i))}{1- \pi(X_i)} +  (\hat \mu_1(X_i) - \hat \mu_0(X_i) ) \Big \} \Big /    \sum_{i=1}^{N} K( \frac{Z_{i} - z}{h})      \\
            {}&  +    O_{\P}\left (  N^{-1/(2+p)} \cdot ( || \hat \mu_0 - \mu_0  ||_{\infty} +  || \hat \mu_1 - \mu_1  ||_{\infty} ) \right )
            \\
          ={}&   \sum_{i=1}^{N} K( \frac{Z_{i} - z}{h}) \Big \{  \dfrac{A_i (Y_i - \mu_1(X_i))}{ \pi(X_i)} - \dfrac{(1-A_i)(Y_i -  \mu_0(X_i))}{1- \pi(X_i)} +  ( \mu_1(X_i) - \mu_0(X_i) ) \Big \} \Big /    \sum_{i=1}^{N} K( \frac{Z_{i} - z}{h})      \\
            {}&  +    O_{\P}\left (  N^{-1/(2+p)} \cdot ( || \hat \mu_0 - \mu_0  ||_{\infty} +  || \hat \mu_1 - \mu_1  ||_{\infty} ) \right ) \\
            {}& +      \sum_{i=1}^{N} K( \frac{Z_{i} - z}{h}) \Big \{  \frac{\pi(X_i) - A_i}{\pi(X_i)}  (\hat \mu_1(X_i) - \mu_1(X_i)) - \frac{ A_i - \pi(X_i) }{1 - \pi(X_i)}  (\hat \mu_0(X_i) - \mu_0(X_i)) \Big \} \Big /    \sum_{i=1}^{N} K( \frac{Z_{i} - z}{h}).
             \end{align*}
By standard analysis of the local constant estimator \cite[see e.g., Section 2 of Li and Racine][]{Li-Racine-2007},   we have
   \[   \sum_{i=1}^{N} K( \frac{Z_{i} - z}{h}) \left (  \frac{A_i Y_i}{ \pi(X_i)} - \frac{(1-A_i)Y_i}{(1- \pi(X_i))}   \right ) \Big /    \sum_{i=1}^{N} K( \frac{Z_{i} - z}{h}) - \tau(z) = O_{\P}(h^2 + 1 / \sqrt{Nh})   \]
and
 \begin{align*}
      &  \sum_{i=1}^{N} K( \frac{Z_{i} - z}{h}) \Big \{  \frac{\pi(X_i) - A_i}{\pi(X_i)}  (\hat \mu_1(X_i) - \mu_1(X_i)) - \frac{ A_i - \pi(X_i) }{1 - \pi(X_i)}  (\hat \mu_0(X_i) - \mu_0(X_i)) \Big \} \Big /    \sum_{i=1}^{N} K( \frac{Z_{i} - z}{h}) \\
      ={}& O_{\P}(h^2 + 1 / \sqrt{Nh}) \cdot ( || \hat \mu_0 - \mu_0  ||_{\infty} +  || \hat \mu_1 - \mu_1  ||_{\infty}).
 \end{align*}
Thus, Theorem 2(a) holds.

In addition, if $|| \hat \mu_a - \mu_a ||_{\infty} = o_{\P}(  N^{-p/(4+2p)} h^{-1/2} )$ for $a = 0, 1$,  then $$ O_{\P}\left (  N^{-1/(2+p)} \cdot ( || \hat \mu_0 - \mu_0  ||_{\infty} +  || \hat \mu_1 - \mu_1  ||_{\infty} ) \right ) = o_{\P}(1/\sqrt{Nh}),$$ then we have
     \begin{align*}
      \hat \tau(z)^{bc}  =&  \sum_{i=1}^{N} K( \frac{Z_{i} - z}{h}) \Big \{  \dfrac{A_i (Y_i - \mu_1(X_i))}{ \pi(X_i)} - \dfrac{(1-A_i)(Y_i -  \mu_0(X_i))}{1- \pi(X_i)} +  ( \mu_1(X_i) - \mu_0(X_i) ) \Big \} \Big /    \sum_{i=1}^{N} K( \frac{Z_{i} - z}{h}) \\ &+ o_{\P}(1/\sqrt{Nh}),
     \end{align*}
           
   and again by the standard analysis of the local constant estimator,
   \[  \sqrt{Nh} \cdot \{ \hat \tau(z)^{bc}  -\tau(z) - \text{bias}( \hat \tau(z)^{bc} ) \} \xrightarrow{d} N(0, \sigma^2), \]
  where $\sigma^2 =  \int K(u)^2 du \cdot  \E[  (\frac{A (Y - \mu_1(X))}{ \pi(X)} - \frac{(1-A)(Y -  \mu_0(X))}{1- \pi(X)} +  ( \mu_1(X) -  \mu_0(X) )   - \tau(z) )^2 \mid Z=z ] \big / f(z)$ and  $\text{bias}( \hat \tau(z)^{bc} ) = O(h^2)$ is a bias term induced by local constant regression.
\end{proof}


\section{\bcol{Additional Simulation Results}}

\subsection{\bcol{MSE for Cases (C1)-(C9)}}

Table \ref{tab-MSE} reports the MSE values of various methods for Cases (C1)–-(C9), corresponding to Figure \ref{fig-MSE} in the manuscript.

\begin{table}[h!]
  \centering
\small
  \caption{Comparison of MSE of various methods for cases (C1)-(C9).}
    \begin{tabular}{cccccccc}
    \toprule
    & & MATCH & MATCH.bc & IPW   & OR    & AIPW  & PSR \\
          Case&       z& \multicolumn{1}{c}{(MSE)} & \multicolumn{1}{c}{(MSE)} & \multicolumn{1}{c}{(MSE)} & \multicolumn{1}{c}{(MSE)} & \multicolumn{1}{c}{(MSE)} & \multicolumn{1}{c}{(MSE)} \\
    \midrule
    & -0.4  & 0.014  & 0.013  & 0.076  & 0.089  & 0.031  & 0.047  \\
& -0.2  & 0.013  & 0.012  & 0.074  & 0.009  & 0.027  & 0.082  \\
C1& 0     & 0.013  & 0.012  & 0.075  & 0.036  & 0.028  & 0.117  \\
& 0.2   & 0.013  & 0.012  & 0.080  & 0.033  & 0.030  & 0.142  \\
& 0.4   & 0.015  & 0.014  & 0.095  & 0.017  & 0.031  & 0.106  \\
    \midrule
    & -0.4  & 0.015  & 0.013  & 0.070  & 0.074  & 0.029  & 0.056  \\
& -0.2  & 0.012  & 0.011  & 0.073  & 0.009  & 0.027  & 0.083  \\
C2& 0     & 0.012  & 0.011  & 0.080  & 0.019  & 0.025  & 0.106  \\
& 0.2   & 0.012  & 0.011  & 0.082  & 0.012  & 0.026  & 0.107  \\
& 0.4   & 0.014  & 0.013  & 0.090  & 0.018  & 0.029  & 0.107  \\
    \midrule
    & -0.4  & 0.015  & 0.014  & 0.074  & 0.018  & 0.029  & 0.098  \\
& -0.2  & 0.013  & 0.012  & 0.078  & 0.014  & 0.028  & 0.065  \\
C3& 0     & 0.013  & 0.012  & 0.088  & 0.050  & 0.028  & 0.059  \\
& 0.2   & 0.014  & 0.012  & 0.092  & 0.031  & 0.030  & 0.068  \\
& 0.4   & 0.015  & 0.014  & 0.104  & 0.127  & 0.030  & 0.215  \\
    \midrule
    & -0.4  & 0.025  & 0.023  & 0.194  & 0.111  & 0.050  & 0.082  \\
& -0.2  & 0.019  & 0.017  & 0.109  & 0.015  & 0.035  & 0.057  \\
C4& 0     & 0.019  & 0.017  & 0.282  & 0.046  & 0.037  & 0.072  \\
& 0.2   & 0.019  & 0.016  & 0.156  & 0.053  & 0.033  & 0.105  \\
& 0.4   & 0.024  & 0.021  & 0.539  & 0.025  & 0.044  & 0.134  \\
    \midrule
    & -0.4  & 0.024  & 0.022  & 0.132  & 0.090  & 0.040  & 0.073  \\
& -0.2  & 0.018  & 0.016  & 0.123  & 0.014  & 0.034  & 0.053  \\
C5& 0     & 0.018  & 0.015  & 0.260  & 0.028  & 0.031  & 0.072  \\
& 0.2   & 0.019  & 0.016  & 0.154  & 0.022  & 0.035  & 0.095  \\
& 0.4   & 0.024  & 0.022  & 0.528  & 0.031  & 0.044  & 0.131  \\
    \midrule
    & -0.4  & 0.024  & 0.022  & 0.170  & 0.026  & 0.044  & 0.099  \\
& -0.2  & 0.021  & 0.018  & 0.138  & 0.029  & 0.035  & 0.046  \\
C6& 0     & 0.018  & 0.016  & 0.419  & 0.068  & 0.035  & 0.048  \\
& 0.2   & 0.019  & 0.017  & 0.185  & 0.042  & 0.033  & 0.056  \\
& 0.4   & 0.024  & 0.021  & 0.511  & 0.130  & 0.050  & 0.151  \\
    \midrule
    & -0.4  & 0.025  & 0.027  & 0.154  & 0.171  & 0.062  & 0.079  \\
& -0.2  & 0.015  & 0.013  & 0.076  & 0.014  & 0.031  & 0.025  \\
C7& 0     & 0.012  & 0.011  & 0.065  & 0.024  & 0.025  & 0.027  \\
& 0.2   & 0.015  & 0.014  & 0.094  & 0.033  & 0.036  & 0.044  \\
& 0.4   & 0.026  & 0.024  & 0.291  & 0.026  & 0.093  & 0.083  \\
    \midrule
    & -0.4  & 0.021  & 0.025  & 0.140  & 0.088  & 0.054  & 0.070  \\
& -0.2  & 0.012  & 0.011  & 0.068  & 0.010  & 0.027  & 0.026  \\
C8& 0     & 0.012  & 0.011  & 0.060  & 0.017  & 0.023  & 0.027  \\
& 0.2   & 0.016  & 0.014  & 0.105  & 0.015  & 0.036  & 0.042  \\
& 0.4   & 0.025  & 0.024  & 0.263  & 0.025  & 0.087  & 0.085  \\
    \midrule
    & -0.4  & 0.024  & 0.021  & 0.143  & 0.132  & 0.061  & 0.061  \\
& -0.2  & 0.013  & 0.012  & 0.073  & 0.021  & 0.029  & 0.028  \\
C9& 0     & 0.013  & 0.012  & 0.068  & 0.025  & 0.023  & 0.023  \\
& 0.2   & 0.017  & 0.015  & 0.104  & 0.043  & 0.035  & 0.042  \\
& 0.4   & 0.027  & 0.026  & 0.286  & 0.063  & 0.087  & 0.089  \\
    \bottomrule
    \end{tabular}
  \label{tab-MSE}
\end{table}

%

\subsection{\bcol{Sensitivity Analysis on Distance Metrics}}
\label{App-C2}

Specifically, we evaluate the performance under three commonly used metrics:
    \begin{itemize}
        \item The Euclidean distance, $$\text{dist}(x, y)= \sqrt{ \sum_{i=1}^p(x_i - y_i)^2}$$ for $x = (x_1, ..., x_p)$ and $y = (y_1, ..., y_p)$.

        \item The Manhattan distance, $$\text{dist}(x, y)= \sum_{i=1}^p |x_i - y_i)|$$ for $x = (x_1, ..., x_p)$ and $y = (y_1, ..., y_p)$.

        \item The Canberra distance, $$\text{dist}(x, y)=  \sum_{i=1}^p \frac{|x_i - y_i|}{ |x_i| + |y_i| }$$ for $x = (x_1, ..., x_p)$ and $y = (y_1, ..., y_p)$.
    \end{itemize}
    The results, presented in Table \ref{tab-comparison1.2} below, show only minor differences in variance and bias, demonstrating that {\bf the proposed estimators perform robustly across various distance metrics.} These findings support the practical flexibility of our method with respect to metric choice.

\begin{table}[h!]
  \centering
 \renewcommand\arraystretch{0.9}
  \small
  \caption{Comparison of Bias and SD of various methods for cases (C1)-(C9) under three commonly used distance metrics,
  with a sample size of $n = 2,000$.}
    \begin{tabular}{cc|cccc|cccc|cccc}
    \toprule
        & & \multicolumn{4}{c|}{\bf Euclidean Distance}  &
         \multicolumn{4}{c|}{\bf Manhattan Distance}
        &   \multicolumn{4}{c}{\bf Canberra Distance} \\
    & & \multicolumn{2}{c}{MATCH} & \multicolumn{2}{c|}{MATCH.bc} & \multicolumn{2}{c}{MATCH} & \multicolumn{2}{c|}{MATCH.bc} & \multicolumn{2}{c}{MATCH} & \multicolumn{2}{c}{MATCH.bc} \\
Case&       z & Bias  & SD    & Bias  & SD    & Bias  & SD    & Bias  & SD    & Bias  & SD    & Bias  & SD \\
\hline
    & -0.4  & -0.009 & 0.119 & -0.008 & 0.113 & -0.015 & 0.122 & -0.014 & 0.115 & -0.010 & 0.122 & -0.009 & 0.117
\\
& -0.2  & 0.007 & 0.114 & 0.007 & 0.108 & 0.006 & 0.110 & 0.005 & 0.106 & 0.012 & 0.115 & 0.012 & 0.109
\\
C1& 0     & 0.017 & 0.111 & 0.015 & 0.106 & 0.008 & 0.109 & 0.008 & 0.104 & 0.010 & 0.115 & 0.010 & 0.109
\\
& 0.2   & 0.020 & 0.112 & 0.019 & 0.108 & 0.009 & 0.108 & 0.008 & 0.103 & 0.011 & 0.115 & 0.011 & 0.109
\\
& 0.4   & -0.005 & 0.123 & -0.005 & 0.117 & -0.003 & 0.118 & -0.004 & 0.113 & 0.001 & 0.122 & -0.001 & 0.117
\\
    \midrule
    & -0.4  & -0.007 & 0.119 & -0.008 & 0.113 & -0.013 & 0.122 & -0.014 & 0.115 & -0.008 & 0.121 & -0.009 & 0.116
\\
& -0.2  & 0.009 & 0.114 & 0.009 & 0.108 & 0.007 & 0.110 & 0.007 & 0.105 & 0.014 & 0.115 & 0.013 & 0.109
\\
C2& 0     & 0.016 & 0.111 & 0.015 & 0.106 & 0.008 & 0.109 & 0.007 & 0.104 & 0.010 & 0.115 & 0.010 & 0.109
\\
& 0.2   & 0.008 & 0.112 & 0.008 & 0.108 & -0.003 & 0.108 & -0.003 & 0.103 & 0.000 & 0.115 & -0.001 & 0.109
\\
& 0.4   & -0.012 & 0.123 & -0.010 & 0.117 & -0.011 & 0.118 & -0.010 & 0.113 & -0.006 & 0.122 & -0.006 & 0.117
\\
    \midrule
    & -0.4  & 0.016 & 0.119 & 0.015 & 0.113 & 0.009 & 0.122 & 0.009 & 0.115 & 0.014 & 0.121 & 0.014 & 0.116
\\
& -0.2  & -0.001 & 0.114 & -0.001 & 0.108 & -0.003 & 0.110 & -0.003 & 0.105 & 0.004 & 0.115 & 0.003 & 0.109
\\
C3& 0     & -0.004 & 0.111 & -0.005 & 0.106 & -0.012 & 0.109 & -0.013 & 0.104 & -0.010 & 0.115 & -0.010 & 0.109
\\
& 0.2   & -0.014 & 0.113 & -0.014 & 0.108 & -0.025 & 0.108 & -0.025 & 0.102 & -0.023 & 0.115 & -0.023 & 0.109
\\
& 0.4   & 0.006 & 0.123 & 0.007 & 0.117 & 0.007 & 0.119 & 0.007 & 0.114 & 0.012 & 0.122 & 0.012 & 0.117
\\
    \midrule
    & -0.4  & -0.042 & 0.148 & -0.027 & 0.143 & -0.038 & 0.149 & -0.024 & 0.146 & -0.042 & 0.148 & -0.026 & 0.144
\\
& -0.2  & 0.007 & 0.140 & 0.014 & 0.130 & 0.006 & 0.138 & 0.012 & 0.130 & 0.002 & 0.134 & 0.010 & 0.126
\\
C4& 0     & 0.017 & 0.139 & 0.015 & 0.130 & 0.008 & 0.132 & 0.006 & 0.123 & 0.016 & 0.140 & 0.016 & 0.131
\\
& 0.2   & 0.036 & 0.132 & 0.026 & 0.125 & 0.023 & 0.136 & 0.014 & 0.127 & 0.025 & 0.134 & 0.017 & 0.127
\\
& 0.4   & 0.031 & 0.155 & 0.014 & 0.150 & 0.037 & 0.152 & 0.020 & 0.148 & 0.034 & 0.148 & 0.017 & 0.144
\\
    \midrule
    & -0.4  & -0.043 & 0.148 & -0.028 & 0.143 & -0.039 & 0.149 & -0.025 & 0.146 & -0.043 & 0.148 & -0.027 & 0.144
\\
& -0.2  & 0.005 & 0.140 & 0.015 & 0.130 & 0.004 & 0.138 & 0.013 & 0.130 & 0.000 & 0.134 & 0.011 & 0.127
\\
C5& 0     & 0.016 & 0.139 & 0.014 & 0.130 & 0.007 & 0.132 & 0.006 & 0.123 & 0.016 & 0.140 & 0.015 & 0.131
\\
& 0.2   & 0.023 & 0.132 & 0.010 & 0.125 & 0.010 & 0.136 & -0.002 & 0.127 & 0.013 & 0.134 & 0.002 & 0.127
\\
& 0.4   & 0.021 & 0.155 & 0.005 & 0.149 & 0.027 & 0.152 & 0.011 & 0.148 & 0.024 & 0.148 & 0.008 & 0.144
\\
    \midrule
    & -0.4  & -0.023 & 0.148 & -0.008 & 0.143 & -0.019 & 0.150 & -0.005 & 0.146 & -0.023 & 0.148 & -0.007 & 0.144
\\
& -0.2  & -0.015 & 0.140 & -0.006 & 0.130 & -0.016 & 0.138 & -0.007 & 0.130 & -0.019 & 0.135 & -0.009 & 0.127
\\
C6& 0     & -0.012 & 0.139 & -0.014 & 0.130 & -0.022 & 0.132 & -0.023 & 0.123 & -0.013 & 0.140 & -0.013 & 0.131
\\
& 0.2   & -0.014 & 0.132 & -0.026 & 0.125 & -0.027 & 0.136 & -0.038 & 0.127 & -0.024 & 0.134 & -0.034 & 0.127
\\
& 0.4   & 0.031 & 0.155 & 0.015 & 0.150 & 0.037 & 0.153 & 0.021 & 0.148 & 0.035 & 0.148 & 0.019 & 0.144
\\
    \midrule
    & -0.4  & -0.064 & 0.150 & -0.086 & 0.146 & -0.073 & 0.145 & -0.094 & 0.141 & -0.056 & 0.146 & -0.078 & 0.143
\\
& -0.2  & -0.024 & 0.114 & -0.019 & 0.109 & -0.028 & 0.112 & -0.022 & 0.108 & -0.023 & 0.114 & -0.017 & 0.109
\\
C7& 0     & -0.016 & 0.108 & -0.007 & 0.104 & -0.022 & 0.110 & -0.013 & 0.105 & -0.023 & 0.109 & -0.014 & 0.104
\\
& 0.2   & -0.009 & 0.121 & -0.001 & 0.116 & -0.012 & 0.124 & -0.003 & 0.119 & -0.013 & 0.123 & -0.005 & 0.118
\\
& 0.4   & 0.008 & 0.164 & -0.006 & 0.160 & 0.019 & 0.156 & 0.008 & 0.153 & 0.011 & 0.165 & -0.004 & 0.162
\\
    \midrule
    & -0.4  & -0.031 & 0.149 & -0.072 & 0.145 & -0.040 & 0.145 & -0.080 & 0.141 & -0.024 & 0.146 & -0.064 & 0.143
\\
& -0.2  & -0.016 & 0.114 & -0.015 & 0.109 & -0.020 & 0.112 & -0.018 & 0.107 & -0.014 & 0.114 & -0.013 & 0.109
\\
C8& 0     & -0.014 & 0.108 & -0.007 & 0.104 & -0.020 & 0.110 & -0.012 & 0.105 & -0.022 & 0.109 & -0.014 & 0.104
\\
& 0.2   & -0.019 & 0.121 & -0.011 & 0.115 & -0.021 & 0.124 & -0.013 & 0.119 & -0.022 & 0.123 & -0.015 & 0.118
\\
& 0.4   & -0.001 & 0.164 & -0.014 & 0.160 & 0.011 & 0.156 & 0.000 & 0.153 & 0.003 & 0.165 & -0.012 & 0.162
\\
    \midrule
    & -0.4  & 0.048 & 0.150 & 0.033 & 0.145 & 0.038 & 0.146 & 0.025 & 0.141 & 0.055 & 0.146 & 0.040 & 0.141
\\
& -0.2  & -0.017 & 0.114 & -0.009 & 0.109 & -0.020 & 0.112 & -0.013 & 0.107 & -0.015 & 0.114 & -0.007 & 0.109
\\
C9& 0     & -0.033 & 0.109 & -0.024 & 0.104 & -0.039 & 0.110 & -0.030 & 0.105 & -0.040 & 0.109 & -0.031 & 0.104
\\
& 0.2   & -0.039 & 0.121 & -0.031 & 0.116 & -0.041 & 0.124 & -0.033 & 0.119 & -0.043 & 0.123 & -0.034 & 0.118
\\
& 0.4   & 0.009 & 0.164 & -0.004 & 0.160 & 0.021 & 0.157 & 0.010 & 0.154 & 0.013 & 0.166 & -0.001 & 0.163
\\
    \bottomrule
    \end{tabular}
  \label{tab-comparison1.2}
\end{table}


\subsection{\bcol{Sensitivity Analysis on a Missing Major Confounder}} \label{App-C3}

\begin{figure}[h]
    \centering
    \includegraphics[width=0.9\linewidth]{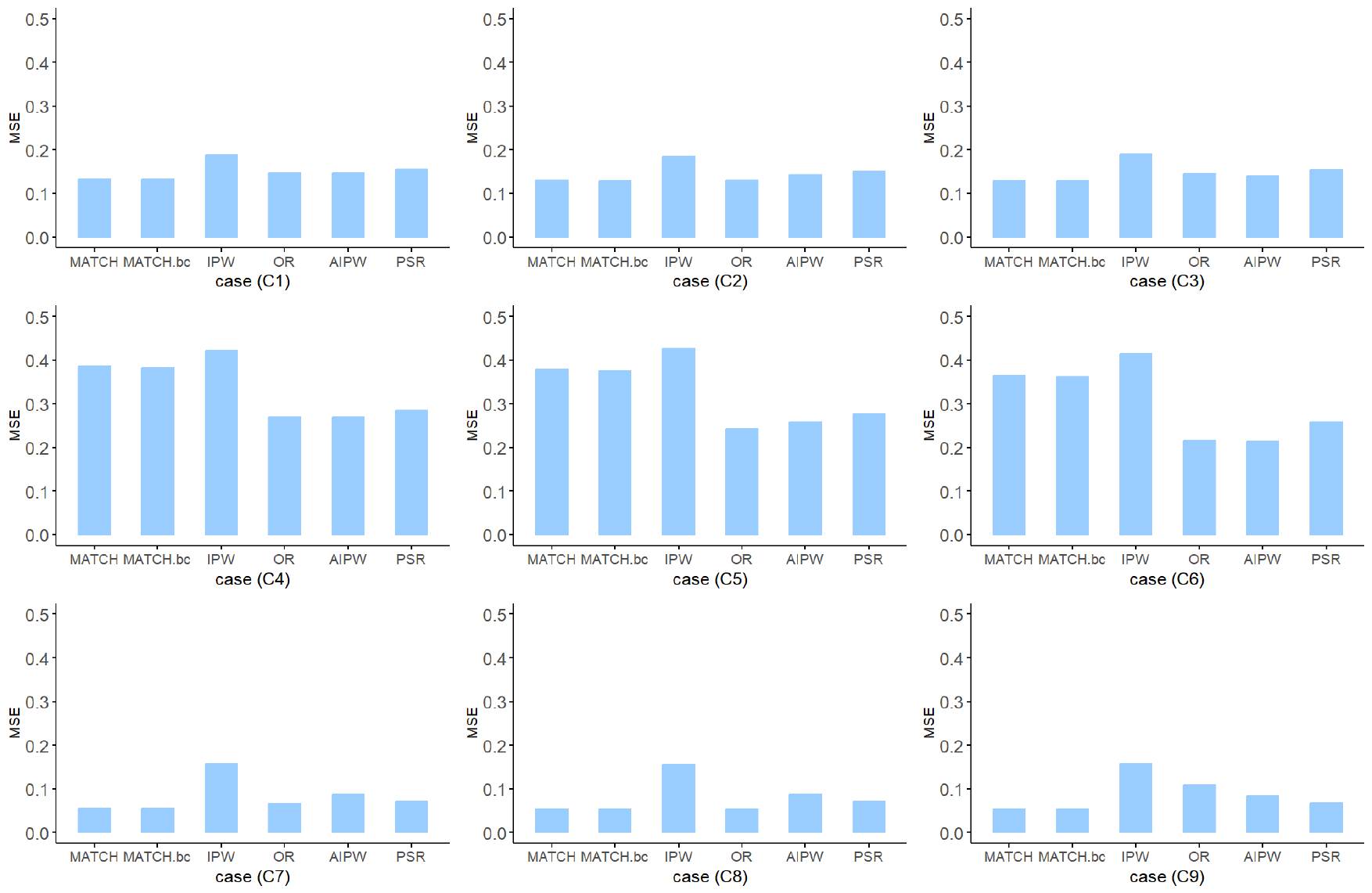}
    \caption{MSE of various methods in cases (C1)--(C9) under omission of a major confounder.}
    \label{fig:MSE_confounder-plots}
\end{figure}

We adopt the data-generating mechanisms of Cases (C1)--(C9), but omit $X_2$
  during estimation by assuming it is unobserved, in order to assess the impact of excluding a key confounder.
  In such cases (we still call them Cases (C1)--(C9)), Figure~\ref{fig:MSE_confounder-plots} shows that the proposed methods still yield lower MSE than the competing approaches.

\subsection{\bcol{Sensitivity Analysis on Number of Matches}}
 \label{App-C4}

We conduct an additional simulation to examine the impact of the number of matches on Bias and SD (standard deviation).
Specifically, let $M$ denote the number of matches per unit. Table \ref{tab-comparison1.4} presents the results for the proposed methods with $M=1, 5,$ and 10, respectively (i.e., One-to-One Matching, One-to-Five Matching, and One-to-Ten Matching). From Table~\ref{tab-comparison1.4}, we observe that:
\begin{itemize}
    \item The Bias remains small and consistent across different numbers of matches;
    \item The SD for \( M = 1 \) is slightly larger than that for \( M = 5 \) or \( M = 10 \). However, the SDs for \( M = 5 \) and \( M = 10 \) are similar.
\end{itemize}
Overall, given that the SD for $M=1$ is only slightly larger than that for $M=5$ and $M=10$, the SD remains relatively stable across different values of \( M \).


\begin{table}[h]
  \centering
 \renewcommand\arraystretch{0.85}
  \small
  \caption{Comparison of Bias and SD of various methods for cases (C1)--(C9) under three numbers of matches,
  with a sample size of $n = 2,000$.}
    \begin{tabular}{cc|cccc|cccc|cccc}
    \toprule
        & & \multicolumn{4}{c|}{\bf One-to-One (M=1)}  &
         \multicolumn{4}{c|}{\bf One-to-Five  (M=5)}
        &   \multicolumn{4}{c}{\bf One-to-Ten (M=10)} \\
   & & \multicolumn{2}{c}{MATCH} & \multicolumn{2}{c|}{MATCH.bc} & \multicolumn{2}{c}{MATCH} & \multicolumn{2}{c|}{MATCH.bc} & \multicolumn{2}{c}{MATCH} & \multicolumn{2}{c}{MATCH.bc} \\
Case&       z & Bias  & SD    & Bias  & SD    & Bias  & SD    & Bias  & SD    & Bias  & SD    & Bias  & SD \\
\hline
    & -0.4  & -0.004 & 0.130 & -0.004 & 0.127 & -0.015 & 0.122 & -0.014 & 0.115 & -0.018 & 0.122 & -0.016 & 0.115
\\
& -0.2  & 0.010 & 0.125 & 0.010 & 0.121 & 0.006 & 0.110 & 0.005 & 0.106 & 0.012 & 0.114 & 0.012 & 0.106
\\
C1& 0     & 0.016 & 0.125 & 0.016 & 0.122 & 0.008 & 0.109 & 0.008 & 0.104 & 0.011 & 0.113 & 0.011 & 0.105
\\
& 0.2   & 0.021 & 0.122 & 0.021 & 0.119 & 0.009 & 0.108 & 0.008 & 0.103 & 0.014 & 0.114 & 0.013 & 0.107
\\
& 0.4   & -0.005 & 0.133 & -0.004 & 0.129 & -0.003 & 0.118 & -0.004 & 0.113 & 0.005 & 0.122 & 0.000 & 0.115
\\
    \midrule
    & -0.4  & -0.010 & 0.134 & -0.010 & 0.131 & -0.008 & 0.122 & -0.009 & 0.116 & -0.013 & 0.124 & -0.014 & 0.115
\\
& -0.2  & 0.007 & 0.121 & 0.007 & 0.119 & 0.016 & 0.109 & 0.015 & 0.104 & 0.010 & 0.112 & 0.010 & 0.104
\\
C2& 0     & 0.009 & 0.120 & 0.010 & 0.118 & 0.008 & 0.111 & 0.007 & 0.105 & 0.010 & 0.108 & 0.010 & 0.101
\\
& 0.2   & 0.002 & 0.125 & 0.002 & 0.122 & 0.000 & 0.110 & 0.000 & 0.105 & 0.003 & 0.109 & 0.002 & 0.101
\\
& 0.4   & -0.018 & 0.129 & -0.017 & 0.127 & -0.012 & 0.120 & -0.012 & 0.114 & -0.003 & 0.121 & -0.005 & 0.114
\\
    \midrule
    & -0.4  & 0.012 & 0.128 & 0.011 & 0.124 & 0.015 & 0.116 & 0.013 & 0.109 & 0.008 & 0.122 & 0.006 & 0.116
\\
& -0.2  & 0.001 & 0.124 & 0.001 & 0.121 & 0.006 & 0.110 & 0.006 & 0.105 & 0.007 & 0.107 & 0.006 & 0.100
\\
C3& 0     & -0.008 & 0.121 & -0.009 & 0.118 & -0.011 & 0.109 & -0.012 & 0.105 & -0.004 & 0.110 & -0.005 & 0.103
\\
& 0.2   & -0.021 & 0.124 & -0.022 & 0.121 & -0.023 & 0.112 & -0.024 & 0.107 & -0.023 & 0.110 & -0.025 & 0.102
\\
& 0.4   & 0.001 & 0.129 & 0.001 & 0.126 & 0.005 & 0.122 & 0.005 & 0.116 & 0.024 & 0.123 & 0.022 & 0.114
\\
    \midrule
    & -0.4  & -0.019 & 0.181 & -0.012 & 0.177 & -0.038 & 0.149 & -0.024 & 0.146 & -0.059 & 0.141 & -0.036 & 0.137
\\
& -0.2  & 0.011 & 0.160 & 0.015 & 0.153 & 0.006 & 0.138 & 0.012 & 0.130 & 0.000 & 0.129 & 0.011 & 0.120
\\
C4& 0     & 0.012 & 0.162 & 0.011 & 0.154 & 0.008 & 0.132 & 0.006 & 0.123 & 0.020 & 0.135 & 0.018 & 0.125
\\
& 0.2   & 0.027 & 0.152 & 0.022 & 0.146 & 0.023 & 0.136 & 0.014 & 0.127 & 0.035 & 0.130 & 0.023 & 0.122
\\
& 0.4   & 0.013 & 0.181 & 0.006 & 0.176 & 0.037 & 0.152 & 0.020 & 0.148 & 0.053 & 0.142 & 0.027 & 0.137
\\
    \midrule
    & -0.4  & -0.024 & 0.181 & -0.016 & 0.177 & -0.039 & 0.149 & -0.024 & 0.145 & -0.056 & 0.146 & -0.035 & 0.141
\\
& -0.2  & 0.009 & 0.159 & 0.014 & 0.153 & 0.003 & 0.134 & 0.013 & 0.126 & -0.006 & 0.139 & 0.009 & 0.128
\\
C5& 0     & 0.012 & 0.154 & 0.013 & 0.147 & 0.012 & 0.132 & 0.010 & 0.123 & 0.019 & 0.130 & 0.017 & 0.119
\\
& 0.2   & 0.008 & 0.152 & 0.001 & 0.145 & 0.021 & 0.134 & 0.008 & 0.126 & 0.023 & 0.129 & 0.006 & 0.120
\\
& 0.4   & -0.004 & 0.182 & -0.011 & 0.177 & 0.024 & 0.152 & 0.008 & 0.148 & 0.038 & 0.145 & 0.011 & 0.140
\\
    \midrule
    & -0.4  & -0.003 & 0.181 & 0.004 & 0.177 & -0.019 & 0.150 & -0.005 & 0.146 & -0.036 & 0.141 & -0.015 & 0.137
\\
& -0.2  & -0.004 & 0.160 & 0.001 & 0.153 & -0.016 & 0.138 & -0.007 & 0.130 & -0.026 & 0.130 & -0.012 & 0.121
\\
C6& 0     & -0.011 & 0.162 & -0.012 & 0.154 & -0.022 & 0.132 & -0.023 & 0.123 & -0.015 & 0.135 & -0.017 & 0.125
\\
& 0.2   & -0.012 & 0.152 & -0.019 & 0.146 & -0.027 & 0.136 & -0.038 & 0.127 & -0.022 & 0.130 & -0.037 & 0.122
\\
& 0.4   & 0.011 & 0.181 & 0.004 & 0.176 & 0.037 & 0.153 & 0.021 & 0.148 & 0.058 & 0.142 & 0.034 & 0.138
\\
    \midrule
    & -0.4  & -0.042 & 0.172 & -0.048 & 0.169 & -0.073 & 0.145 & -0.094 & 0.141 & -0.072 & 0.139 & -0.113 & 0.137
\\
& -0.2  & -0.008 & 0.128 & -0.005 & 0.124 & -0.028 & 0.112 & -0.022 & 0.108 & -0.030 & 0.114 & -0.023 & 0.106
\\
C7& 0     & -0.004 & 0.121 & 0.001 & 0.117 & -0.022 & 0.110 & -0.013 & 0.105 & -0.028 & 0.108 & -0.016 & 0.102
\\
& 0.2   & -0.005 & 0.137 & 0.001 & 0.134 & -0.012 & 0.124 & -0.003 & 0.119 & -0.010 & 0.121 & -0.002 & 0.114
\\
& 0.4   & -0.018 & 0.197 & -0.019 & 0.191 & 0.019 & 0.156 & 0.008 & 0.153 & 0.038 & 0.153 & 0.008 & 0.152
\\
    \midrule
    & -0.4  & -0.026 & 0.162 & -0.040 & 0.158 & -0.033 & 0.141 & -0.074 & 0.138 & -0.020 & 0.132 & -0.091 & 0.129
\\
& -0.2  & -0.009 & 0.125 & -0.006 & 0.122 & -0.012 & 0.115 & -0.011 & 0.110 & -0.018 & 0.112 & -0.019 & 0.105
\\
C8& 0     & -0.009 & 0.120 & -0.004 & 0.117 & -0.020 & 0.108 & -0.013 & 0.103 & -0.023 & 0.109 & -0.012 & 0.103
\\
& 0.2   & -0.022 & 0.142 & -0.017 & 0.139 & -0.022 & 0.123 & -0.014 & 0.118 & -0.020 & 0.117 & -0.011 & 0.112
\\
& 0.4   & -0.028 & 0.204 & -0.029 & 0.198 & 0.011 & 0.163 & -0.004 & 0.161 & 0.031 & 0.153 & 0.003 & 0.152
\\
    \midrule
    & -0.4  & 0.012 & 0.171 & 0.010 & 0.168 & 0.038 & 0.146 & 0.025 & 0.141 & 0.093 & 0.138 & 0.064 & 0.134
\\
& -0.2  & -0.009 & 0.128 & -0.005 & 0.124 & -0.020 & 0.112 & -0.013 & 0.107 & -0.012 & 0.114 & -0.002 & 0.106
\\
C9& 0     & -0.020 & 0.121 & -0.015 & 0.118 & -0.039 & 0.110 & -0.030 & 0.105 & -0.046 & 0.108 & -0.033 & 0.102
\\
& 0.2   & -0.032 & 0.138 & -0.027 & 0.134 & -0.041 & 0.124 & -0.033 & 0.119 & -0.041 & 0.121 & -0.033 & 0.114
\\
& 0.4   & -0.017 & 0.197 & -0.018 & 0.191 & 0.021 & 0.157 & 0.010 & 0.154 & 0.041 & 0.153 & 0.011 & 0.152
\\
    \bottomrule
    \end{tabular}
  \label{tab-comparison1.4}
\end{table}

\newpage
\subsection{\bcol{Additional Simulation In the Presence of Extreme Estimated Propensity Scores}}\label{App-C5}

Below, we present empirical simulations to evaluate the robustness of the proposed methods and propensity weighting-based methods under practical (and near) violations of the overlap assumption. We consider three data-generating mechanisms (C10)--(C12) as follows:

{\bf (C10)} $\P(A=1\mid X) = \{ 1 + \exp( -(X_1^2, X_2^2, X_3^2) \bm{\alpha}) \}^{-1}$, where $X_1 \sim \text{Unif}(-1/2, 1/2)$, $X_2$ takes values of $\{0, 1, 2\}$ with equal probabilities, $X_3 \sim N(0, 1)$, and $\bm{\alpha} =(8, 1/2, -5/4)$.
      The potential outcomes $Y(1)$ and $Y(0)$ are generated the same as Case (C4) in the manuscript.

       {\bf(C11)} $(X, A)$ are generated the same as Case (C10),
      $Y(1)$ and $Y(0)$ are generated the same as Case (C5) in the manuscript.

       {\bf (C12)} $(X, A)$ are generated the same as Case (C10), $Y(1)$ and $Y(0)$ are generated the same as Case (C6) in the manuscript.

In Cases (C10)--(C12), several true propensity scores are very close to 1. For each simulation with a sample size of $n=2,000$, the minimum true propensity score is approximately $10^{-7}$, and about 2\% of the true propensity scores are on the order of $10^{-3}$.  In fact, the data-generating mechanisms for Cases (C10)--(C12) are identical to those in Cases (C4)--(C6). However, to ensure that the estimated propensity scores are equal to or close to 0 or 1, we adopt a slightly different method for estimating the propensity scores.
 \begin{itemize}
     \item For cases (C4)--(C6) in the manuscript, the propensity score model is specified as a logistic function of $X=(X_1, X_2, X_3)$ with main effects.
     The propensity score model is misspecified, and the estimated propensity scores do not approach 0 or 1.
     \item For cases (C10)--(C12) in this response, the propensity score model is specified as a logistic function of $(X_1^2, X_2^2, X_3^2)$. The propensity score model is correctly specified, and several estimated propensity scores are extremely close to zero, and in some cases, even exactly zero.
 \end{itemize}
Since a different method is used to estimate the propensity scores, we use ``cases (C10)--(C12)" rather than referring to them as ``cases (C4)--(C6)".

\bcol{\begin{table}[h!]
  \centering
  \setlength{\tabcolsep}{1.7mm}
  \renewcommand\arraystretch{1.2}
  \small
  \caption{Comparison of Bias and SD of various methods for cases (C10)-(C12), with a sample size of $n = 2,000$.}
    \begin{tabular}{cccccccccccccc}
    \toprule
    & & \multicolumn{2}{c}{MATCH} & \multicolumn{2}{c}{MATCH.bc} & \multicolumn{2}{c}{IPW} & \multicolumn{2}{c}{OR} & \multicolumn{2}{c}{AIPW} & \multicolumn{2}{c}{PSR} \\
Case&       z& Bias  & SD    & Bias  & SD    & Bias  & SD    & Bias  & SD    & Bias  & SD    & Bias  & SD \\
\hline
    & -0.4  & -0.042 & {\bf 0.148} & -0.027 & {\bf 0.143} & -0.089 & {\bf 2.186} & -0.288 & 0.153 & -0.076 & {\bf 1.740} & 0.062 & 0.223
\\
& -0.2  & 0.007 & {\bf 0.140} & 0.014 & {\bf 0.130} & 0.116 & {\bf 9.485} & 0.031 & 0.118 & 0.081 & {\bf 6.585} & 0.088 & 0.192
\\
C10 & 0     & 0.017 & {\bf 0.139} & 0.015 & {\bf 0.130} & -0.213 & {\bf 5.962} & 0.194 & 0.103 & -0.142 & {\bf 4.337} & 0.096 & 0.186
\\
& 0.2   & 0.036 & {\bf 0.132} & 0.026 & {\bf 0.125} & 0.030 & {\bf 1.040} & 0.197 & 0.112 & 0.032 & {\bf 0.655} & 0.142 & 0.188
\\
& 0.4   & 0.031 & {\bf 0.155} & 0.014 & {\bf 0.150} & -0.129 & {\bf 4.337} & 0.034 & 0.153 & -0.104 & {\bf 3.322} & 0.201 & 0.244
\\
    \midrule
    & -0.4  & -0.043 & {\bf 0.148} & -0.028 & {\bf 0.143} & -0.090 & {\bf 2.235} & -0.251 & 0.153 & -0.076 & {\bf 1.743} & 0.084 & 0.223
\\
& -0.2  & 0.005 & {\bf 0.140} & 0.015 & {\bf 0.130} & 0.114 & {\bf 9.426} & 0.025 & 0.117 & 0.083 & {\bf 6.585} & 0.094 & 0.192
\\
C11 & 0     & 0.016 & {\bf 0.139} & 0.014 & {\bf 0.130} & -0.215 & {\bf 5.999} & 0.132 & 0.103 & -0.142 & {\bf 4.326} & 0.094 & 0.186
\\
& 0.2   & 0.023 & {\bf 0.132} & 0.010 & {\bf 0.125} & 0.020 & {\bf 1.044} & 0.093 & 0.111 & 0.022 & {\bf 0.655} & 0.110 & 0.188
\\
& 0.4   & 0.021 & {\bf 0.155} & 0.005 & {\bf 0.149} & -0.135 & {\bf 4.316} & 0.078 & 0.153 & -0.113 & {\bf 3.340} & 0.194 & 0.244
\\
    \midrule
    & -0.4  & -0.023 & {\bf 0.148} & -0.008 & {\bf 0.143} & -0.072 & {\bf 2.202} & -0.008 & 0.154 & -0.059 & {\bf 1.770} & 0.150 & 0.223
\\
& -0.2  & -0.015 & {\bf 0.140} & -0.006 & {\bf 0.130} & 0.120 & {\bf 9.667} & -0.121 & 0.118 & 0.079 & {\bf 6.659} & 0.069 & 0.193
\\
C12 & 0     & -0.012 & {\bf 0.139} & -0.014 & {\bf 0.130} & -0.220 & {\bf 5.715} & -0.241 & 0.103 & -0.153 & {\bf 4.197} & 0.037 & 0.186
\\
& 0.2   & -0.014 & {\bf 0.132} & -0.026 & {\bf 0.125} & -0.005 & {\bf 1.061} & -0.165 & 0.111 & 0.000 & {\bf 0.658} & 0.051 & 0.188
\\
& 0.4   & 0.031 & {\bf 0.155} & 0.015 & {\bf 0.150} & -0.124 & {\bf 4.335} & 0.323 & 0.153 & -0.104 & {\bf 3.405} & 0.276 & 0.244
\\
    \bottomrule
    \end{tabular}
  \label{tab-comparison1.1}
\end{table}}

\section{\bcol{Description of Variables for Application}}

Table \ref{tab-description} summarizes the variables used in Application of the main text.

 \begin{table}[h]
\caption{Descriptions of variables}
\centering
\footnotesize
 \label{tab-description}
\begin{tabular}{llll}
		\hline
	 {\bf Variable}  &  {\bf Name}   & {\bf Proportion/Mean}  &   {\bf (Q1, Q3)}  \\
		\hline			
		Outcome      &  PASI 75  &       &    \\
						   &      \qquad  yes (\%)        &  46.80\%  \\
						   &      \qquad  no (\%)       &  53.20\%  \\
		Treatment    &  Biologics  &   \\
							&      \qquad  yes (\%)       &  30.05\%  \\
							&      \qquad  no  (\%)      &  69.95\%  \\   \hline 
		\multirow{13}{*}{Covariates}  & Baseline PASI  & 11.99  & (3.60, 17.83)  \\
		&   Baseline BSA  &  20.35 & (4.38, 30.00)   \\
		&  Baseline DLQI & 8.44 & (3.00, 12.00)   \\
		&  Age &   41.70  &  (30.00, 53.00)  \\
		&  BMI &   24.45 & (21.77, 26.12)    \\
		&     Employment  \\
				&      \qquad  Full-time (\%)        &  61.29\%  \\
					&      \qquad  Part-time (\%)        &  38.71\%  \\
		&  Marital status    &\\
				&      \qquad  Married (\%)       &  74.24\%  \\
					&      \qquad  Unmarried (\%)        & 25.76\%  \\
		&  Education &  \\
		&      \qquad  College and higher (\%)       &  28.31\%  \\
					&      \qquad  High school and lower (\%)        &  71.69\%  \\
		&  Insurance &  \\
		&      \qquad  Free or commercial  medical care (\%)       &  8.28\%  \\
					&      \qquad  General government funded medical care (\%)        &  91.72\%  \\
		&   Nail involvement  &    \\
			 &      \qquad  yes (\%)      &  6.75\%  \\
					&      \qquad  no (\%)        &  93.25\%  \\
		 &  Sex     &    \\
		 			 &      \qquad  Female (\%)      &  35.23\%  \\
					&      \qquad  Male (\%)        &  64.77\%  \\
		 &  Smoking   &   \\
				 &      \qquad  Ex-smokers (\%)      &  5.60\%  \\
					&      \qquad  Current smokers (\%)       &  27.46\%  \\
					&      \qquad  Non-smokers (\%)        &  66.94\%  \\
		 & Comorbidity conditions &  \\
		 		 &      \qquad  No comorbidity (\%) &  78.27\%  \\
					&      \qquad  Have comorbidity (\%) &  12.39\%  \\
					&      \qquad  Not clear (\%)       &  9.34\%  \\
		 &   \multicolumn{3}{l}{All the interactions between BMI and all other covariates.}  \\
		 &   \multicolumn{3}{l}{All the interactions between Age and all other covariates.}  \\
		 &   \multicolumn{3}{l}{All the interactions between Baseline PASI and all other covariates.}  \\
		 &   \multicolumn{3}{l}{All the interactions between Baseline DLQI and all other covariates.}  \\
		 &   \multicolumn{3}{l}{All the interactions between Baseline BSA and all other covariates.}  \\
		\hline	
	\end{tabular}
		\vskip 0.6cm
	\begin{flushleft}
	Note: DLQI and Baseline BSA denote self-reported Dermatology Life Quality Index and  baseline Body Surface Area, respectively;  Nail involvement refers to whether more than 5 nails are seriously affected.
	\end{flushleft}
\end{table}
\normalsize

\end{document}